\documentclass[superscriptaddress,aps,journal=prl,preprint]{revtex4-2}
\usepackage{graphicx}
\usepackage{epstopdf}
\usepackage{bm,amsmath,amssymb} % for math
\usepackage{color, soul} % for highlight
\usepackage{dcolumn}   % needed for some tables
\usepackage{multirow}
\usepackage[colorlinks=true, linkcolor=black, citecolor=black, urlcolor=black]{hyperref}
\newcommand{\etal}{{\em et al}.\ }
 
\newcommand{\note}[1]{{\color{black}{{#1}}}}

\usepackage{lineno}
\begin{document}
%\linenumbers
\title{Giant piezoelectric effects of topological structures in stretched ferroelectric membranes}
\author{Yihao Hu}
\affiliation{Key Laboratory for Quantum Materials of Zhejiang Province, Department of Physics, School of Science, Westlake University, Hangzhou, Zhejiang 310024, China}
\author{Jiyuan Yang}
\affiliation{Key Laboratory for Quantum Materials of Zhejiang Province, Department of Physics, School of Science, Westlake University, Hangzhou, Zhejiang 310024, China}
\author{Shi Liu}
\email{liushi@westlake.edu.cn}
\affiliation{Key Laboratory for Quantum Materials of Zhejiang Province, Department of Physics, School of Science, Westlake University, Hangzhou, Zhejiang 310024, China}
\affiliation{Institute of Natural Sciences, Westlake Institute for Advanced Study, Hangzhou, Zhejiang 310024, China}

\begin{abstract}
Freestanding ferroelectric oxide membranes emerge as a promising platform for exploring the interplay between topological polar ordering and dipolar interactions that are continuously tunable by strain. Our  investigations combining density functional theory (DFT) and deep-learning-assisted molecular dynamics simulations demonstrate that DFT-predicted strain-driven morphotropic phase boundary involving monoclinic phases manifest as diverse domain structures at room temperatures, featuring continuous distributions of dipole orientations and mobile domain walls. 
Detailed analysis of dynamic structures reveals that the enhanced piezoelectric response observed in stretched PbTiO$_3$ membranes results from small-angle rotations of dipoles at domain walls, distinct from conventional polarization rotation mechanism and adaptive phase theory inferred from static structures. We identify a ferroelectric topological structure, termed ``dipole spiral," which exhibits a giant intrinsic piezoelectric response ($>$320 pC/N). This helical structure, possessing a rotational zero-energy mode, unlocks new possibilities for exploring chiral phonon dynamics and dipolar Dzyaloshinskii-Moriya-like interactions.
\end{abstract}

\maketitle
\newpage
% thin film epitaxy --> freestanding membrance
% PTO map out the phase diagram
%\section{Introduction}
\date{\today}% It is always \today, today,

Strain engineering of ferroelectric oxides through thin film epitaxy has greatly advanced the understanding of ferroelectric physics and led to the realization of novel topological polar structures and functionalities~\cite{Tang15p547,Yadav16p198,Das19p368}. By exploiting the lattice mismatch between ferroelectric oxides and their substrates, the interactions among spin, charge, orbital, lattice, and domain degrees of freedom can be deterministically controlled~\cite{Ramesh19p257}. 
Nevertheless, the effectiveness of epitaxial strain is generally limited to $\approx$2\%.  Beyond this threshold, defects and dislocations tend to form at the ferroelectric-substrate interface, leading to strain relaxation~\cite{Fernandez22p1521}. The number of strain states for a ferroelectric oxide is further restricted by the availability of high-quality substrates. Recent advancements in synthesizing single-crystal, freestanding oxide membranes have opened new avenues~\cite{Lu16p1255,Dong19p475,Xu20p3141}, enabling strain states up to an unprecedented level ($\approx$8\%)~\cite{Hong20p71,Cai22p13} and integration with silicon-based technologies~\cite{Han22p63, Huang22p262}. 
Moreover, the freestanding membrane, adaptable to continuously variable isotropic and anisotropic strains~\cite{Xu20p3141}, allows for in-depth investigations into the intricate interplay between topological polar ordering and dipole correlations. A general approach to predicting the strain phase diagram under experimental conditions will facilitate the discovery of novel emergent phenomena in ferroelectric membranes. The challenge is to bridge the gap between zero-Kelvin, first-principles-based, unit-cell-level calculations and measurable macroscopic properties, which are often significantly influenced by mesoscopic domain structures.
 
Pertsev~\etal~pioneered the mapping of ferroelectric perovskite structures against temperature and misfit strain using Landau-Devonshire theory based on empirical thermodynamic potentials~\cite{Pertsev03p054107}. Dieguez~\etal~subsequently demonstrated that predictions from this method are sensitive to parameters fitted to experimental data, highlighting the importance of an {\em ab initio} approach~\cite{Diguez05p144101}. Although first-principles density functional theory (DFT) is commonly used to predict phase diagrams~\cite{Dieguez10p094105,Angsten17p174110}, the single-domain approximation introduced to reduce computational costs neglects the impacts of domain structures. In contrast, phase-field methods, effective in predicting three-dimensional (3D) domain structures, rely heavily on empirical parameters and lack atomic-level details. Here, 
we employ deep potential molecular dynamics (DPMD)~\cite{Zhang18p143001} simulations to construct phase diagrams at finite temperatures, advancing beyond the single-domain assumption. 
%These simulations, utilizing a deep neural network-based force field trained exclusively with first-principles data~\cite{Wu23p144102}, demonstrate a fully {\em ab initio} approach to temperature-dependent macroscopic domain properties while providing atomistic insights. 

Taking PbTiO$_3$ membranes for example, we show that while DFT calculations indicate a tensile-strain-driven morphotropic phase boundary (MPB) with competing phases~\cite{Jaffe54p809}, this feature becomes absent in thermally active environments. Instead, the flat potential energy landscape results in diverse domain structures with flexible dipoles and mobile domain walls. 
DPMD simulations reveal that the dynamic structure of the $c/a$ two-domain state exhibits a broad and continuous distribution of dipole orientations.
The collective and coordinated small-angle rotations of dipoles at domain walls underlie the enhanced piezoelectric strain coefficient ($d_{33}$) observed experimentally in stretched PbTiO$_3$ membranes~\cite{Han23p2808}, distinct from conventional polarization rotation mechanism~\cite{Fu00p281,Kutnjak06p956} and adaptive phase theory~\cite{Jin03p3629,Jin03p197601}. Interestingly, further stretching the membrane could activate spontaneous and stochastic oscillations of 90$^\circ$ domain walls, leading to an even higher $d_{33}$ value of $\approx$250 pC/N, three times that of a single domain ($\approx$80 pC/N).
We further discover a ferroelectric topological structure, the dipole spiral, characterized by canted dipoles that progressively rotate around the out-of-plane direction. This helical dipolar structure supports a giant piezoelectric response ($>320$ pC/N) through small-angle dipole rotations.

We start by constructing the strain phase diagram for PbTiO$_3$ across a wide range of tensile strains, based on high-throughput DFT calculations. These calculations serve as a mean-field-like analysis for energy variation with respect to polarization ($P$) orientation. 
All first-principles calculations are performed with the projector augmented-wave (PAW) method \cite{Blochl94p17953,Kresse99p1758}, using the Vienna \textit{ab initio} simulation package (\texttt{VASP}) \cite{Kresse96p11169,Kresse96p15}. The exchange-correlation functional is treated within the generalized gradient approximation of Perdew-Burke-Ernzerhof revised for solids (PBEsol) type~\cite{Perdew08p136406}.
For a given strain state, the in-plane lattice parameters ($a_{\rm IP}$ and $b_{\rm IP}$) of a five-atom unit cell are fixed, while the atomic coordinates and out-of-plane lattice constant are fully optimized. 
\note{This setup closely resembles the application of orthogonal strains to freestanding membranes, which is a common  scenario  in experimental settings~\cite{Xu20p3141,Hong20p71,Han23p2808}.}
To access competing polar states, multiple initial configurations with polarization pointing in different directions are used.
A kinetic energy cutoff of 800 eV, a $k$-point spacing of 0.3 \AA$^{-1}$ for the Brillouin zone integration, and a force convergence threshold of 0.001 eV/\AA~are used to converge the energy and atomic forces.

\begin{figure}
	\begin{center}
		\includegraphics[width=0.8\textwidth]{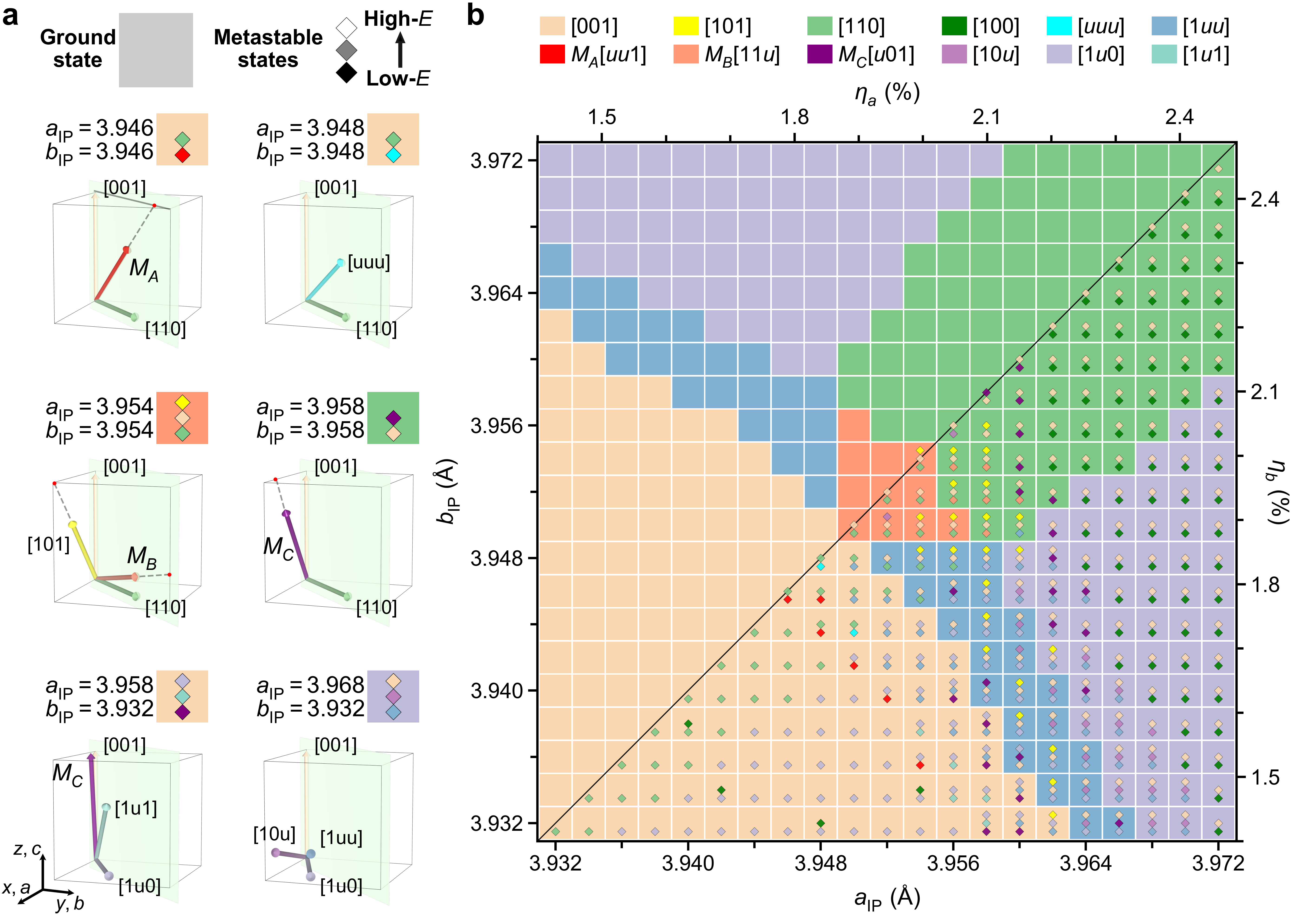}
	\end{center}
	\caption{\textbf{DFT strain multiphase diagram for PbTiO$_3$ membranes.} \textbf{a,} Unique polar states stabilized by tensile strains. A plaquette in the phase diagram encodes all possible phases that a five-atom unit cell can sustain under a specific strain condition. The square's background color corresponds to the ground state, while other metastable phases are indicated by markers arranged vertically by their energies ($E$). \textbf{b,} Phase diagram illustrating the competitions among phases with comparable energies. Considering the exchange symmetry between in-plane $a$ and $b$ axes, the phase compositions are explicitly depicted only within the bottom right triangular region. The strain $\eta$ is computed relative to the DFT ground-state value ($a_0=b_0=3.877$~\AA).
 \label{DFT}}
\end{figure}

We introduce a ``multiphase" diagram to illustrate the competitions among phases with comparable energies (within 6 meV/atom). 
Twelve unique polar states (see Fig.~\ref{DFT}\textbf{a}) are identified, each categorized by the polarization direction while considering the exchange symmetry between in-plane $a$ and $b$ axes.
For strains close to equal-biaxial conditions ($a_{\rm IP}=b_{\rm IP}$), we observe some well-known phases: a tetragonal ($T$) phase with its polarization along the pseudocubic [001] axis; orthorhombic ($O$) [110] and [101] phases with polarization along the face diagonal directions; and a rhombohedral ($R$, denoted as $[111]$) phase with nearly equal magnitudes of $P_x$, $P_y$, and $P_z$. There are also three monoclinic phases introduced by Vanderbilt and Cohen~\cite{Vanderbilt01p094108}: $M_A$ with $P_x\approx P_y < P_z$ (denoted as $[uu1]$ with $u<1$), serving as a bridge between the [001] and [111] phases; $M_B$ with $P_x\approx P_y > P_z$ (denoted as $[11u]$), which connects the [110] and [111] phases; and $M_C$ with a space group of $Pm$ (denoted as $[u01]$), intermediate between the [001] and [101] phases. A strongly anisotropic biaxial strain induce four additional phases: two distorted $R$ phases, $[1uu]$ with $P_x > P_y \approx P_z$ and $[1u1]$ with $P_x \approx P_z > P_y$, and two distorted $O$ phases, $[1u0]$ and $[10u]$.
Finally, under a sufficiently large tensile strain along the $a$ axis, the [100] state becomes competitive. 
It is evident from Fig.~\ref{DFT}\textbf{b} that a variety of strain conditions can stabilize multiple phases. For example, at $a_{\rm IP}=b_{\rm IP}=3.946$~\AA, the energies of [001], $M_A$, and [110] phases are within 1 meV/atom.  
%The anisotropic strain specified by $a_{\rm IP}=3.968$~\AA~and $b=3.932$~\AA~results in the coexistence of $[1u0]$, $[1uu]$, $[10u]$, and $[001]$ states. 

The multiphase diagram suggests that a tensile in-plane strain leads to a flat potential energy landscape with respect to polarization rotation in PbTiO$_3$, a hallmark of MPB ~\cite{Ahart08p545}. The emergence of phase competitions involving various $M$ phases supports a $M$-phase-mediated polarization rotation mechanism~\cite{Guo2000p5423,Noheda99p2059}. The question is whether this MPB-like feature persists at finite temperatures. To address this, we perform large-scale DPMD simulations to investigate the polar ordering at finite temperatures. For simplicity, we focus on isotropic in-plane strains ($a_{\rm IP}=b_{\rm IP}$). 
The DP model can reproduce various properties of PbTiO$_3$, including phonon spectra of tetragonal and cubic phases, the temperature-driven phase transition, and topological textures such as polar vortex lattice in PbTiO$_3$/SrTiO$_3$ superlattices~\cite{Wu23p144102}. We have developed an online 
{\href{https://github.com/huiihao/Spiral}{{notebook}}}~\cite{datayhh}
on Github to share the training database, force field model, training metadata, and essential input and output files.
Further details about MD simulations using \texttt{LAMMPS} \cite{Plimpton95p1} can be found in Supplementary Material~\cite{SIyhh}.

\begin{figure}
	\begin{center}
		\includegraphics[width=1\textwidth]{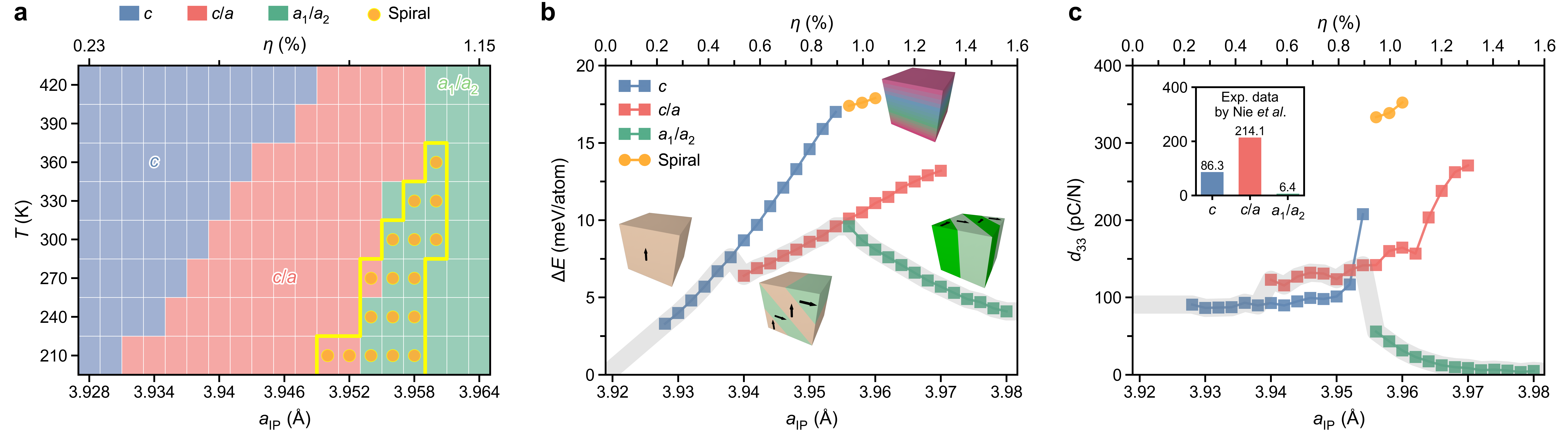}
	\end{center}
	\caption{{\bf Thermodynamic stability and piezoelectric response of domain structures.} \textbf{a,} MD strain-temperature domain stability diagram. The yellow-colored boundary highlights the strain states supporting dipole spirals. The strain $\eta$ is computed in reference to the MD ground-state value at 300~K ($a_0=b_0=3.919$~\AA).
 \textbf{b,} Relative thermodynamic stability and  \textbf{c,} piezoelectric coefficients of different domain structures with respect to isotropic in-plane strains at 300~K. The thick shaded line traces the most stable domain structure. The inset reports experimental $d_{33}$ values of PbTiO$_3$ membranes~\cite{Han23p2808}.
 \label{MD}}
\end{figure}

The strain-temperature domain stability diagram 
is presented in Fig.~\ref{MD}\textbf{a}, revealing three well-known domain structures and a novel, metastable topological structure that resembles a spin spiral~\cite{Tsunoda89p10427,Tsunoda93p133}. The three recognized domain morphologies are: a single $c$-domain state comprised solely of [001] domains; a $c/a$ two-domain state with [001] and [100] (or [010]) domains; and an $a_1/a_2$ two-domain state with [100] and [010] domains. Predicted without empirical parameters, the strain-driven evolution of these domain structures, $c\rightarrow c/a\rightarrow a_1/a_2$, agrees well with results from phase-field simulations~\cite{Li01p3878,Kavle22p2203469}.
The topological structure, which we name ``dipole spiral," features canted dipoles that progressively rotate around the [001] direction (see detailed discussions below). 
Figure~\ref{MD}\textbf{b} plots the energies of domain structures at 300~K against $a_{\rm IP}$. 
For certain strains \note{($\eta$, computed relative to the ground-state in-plane lattice constant of the $c$-domain PbTiO$_3$ at 300~K)}, multiple domain morphologies can coexist. For example, in the range of \note{$0.94\%<\eta<1.05\%$}, $a_1/a_2$, $c/a$, and the dipole spiral are all stable in MD simulations.
The dipole spiral, albeit energetically higher than the lowest-energy state, demonstrates significant robustness across a broad temperature and strain range (Fig.~\ref{MD}\textbf{a}).
\note{It is noted that we observe a discontinuous evolution from a $c$-domain state to a dipole spiral at a critical strain of $a_{\rm IP}\approx3.954$~\AA~(see Fig.~S12).}
These findings convey that MPB-like phase competitions, predicted by zero-Kelvin DFT calculations, actually manifest as complex domain structures at finite temperatures, prompting an essential inquiry: can the domain structure enhance the piezoelectric response in the absence of MPB?

Using finite-field MD simulations, we quantify $d_{33}$ of stretched membranes via the direct piezoelectric effect, $[\partial{\eta}_3/\partial{\mathcal{E}_3}]|_{\sigma_3=0}$, where ${\eta}_3$ is the strain change along the $z$ aixs due to an out-of-plane electric field ($\mathcal{E}_3$). Figure~\ref{MD}\textbf{c} shows that
the $c/a$ domain structure yields larger $d_{33}$ values than the single $c$-domain under the same strain conditions ($3.94<a_{\rm IP}<3.955$~\AA), indicating 90$^{\circ}$ domain walls enhance the piezoelectric response. For higher tensile strains ($a_{\rm IP}>3.955$~\AA) which favor the $a_1/a_2$ state, $d_{33}$ diminishes rapidly due to minimal out-of-plane polarization. The concave characteristic of the $d_{33}$ versus $a_{\rm IP}$ curve, highlighted by the thick shaded line in Fig.~\ref{MD}\textbf{c}, agrees quantitatively with the trend observed in experiments with freestanding PbTiO$_3$ membranes~\cite{Han23p2808}. Notably, the $d_{33}$ value of the $c/a$ state experiences a jump when $a_{\rm IP}$ is beyond a critical value of  3.962~\AA, surpassing 250 pC/N and significantly exceeding the bulk value of $\approx$80 pC/N. 

\begin{figure}
	\begin{center}
		\includegraphics[width=1\textwidth]{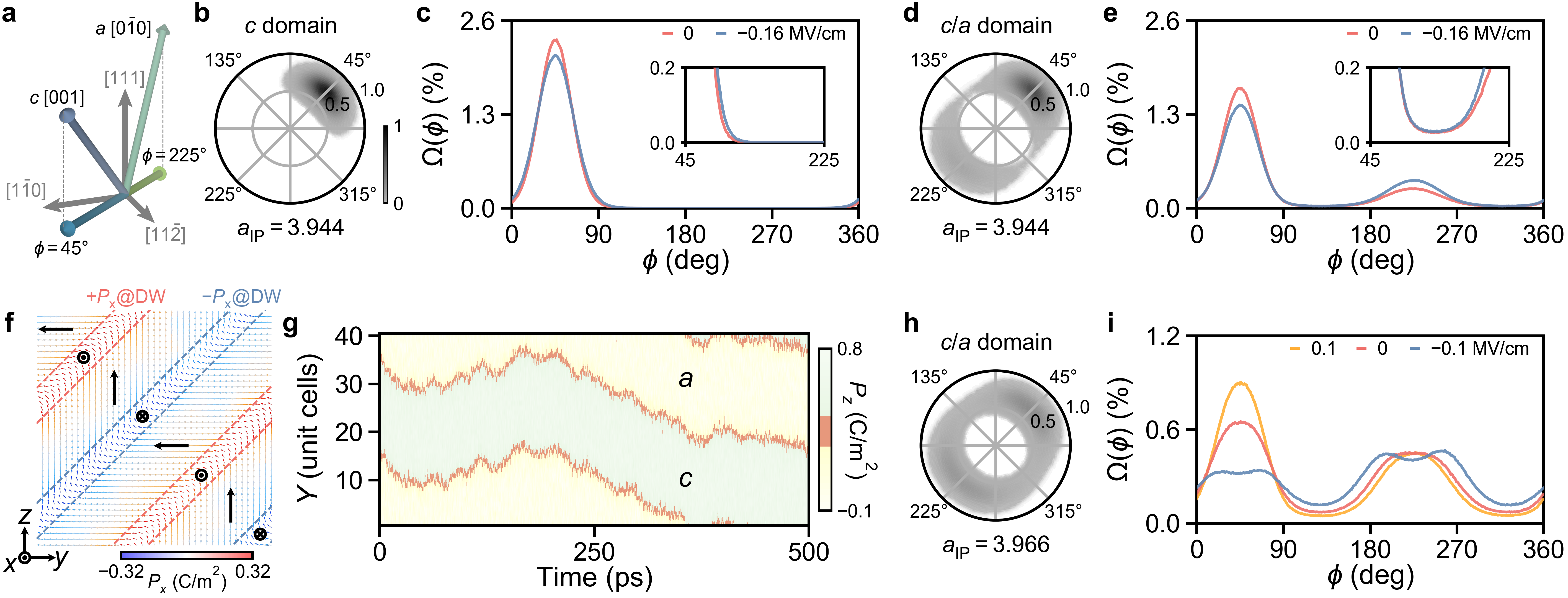}
	\end{center}
	\caption{\textbf{Enhanced piezoelectricity in stretched PbTiO$_3$ membranes with $c/a$ two-domain states.} \textbf{a,} Schematic illustration of a $[0\bar{1}0]$ dipole in the $a$ domain and a $[001]$ dipole in the $c$ domain projected onto the \{111\} plane. Dipole orientation distributions in \textbf{b-c,} single-$c$ domain, and \textbf{d-e,} $c/a$ two-domain states under the same strain condition ($a_{\rm IP}=3.944$~\AA). The distributions are plotted in polar coordinates viewed along [111] in (\textbf{b}) and (\textbf{d}). The distributions of azimuthal angles ($\phi$) in the \{111\} plane and their changes to an out-of-plane field ($\mathcal{E}_3$) are presented in (\textbf{c}) and (\textbf{e}), with insets providing zoomed-in views. \textbf{f-i,} $c/a$ domain structures in strongly stretched membranes ($a_{\rm IP}=3.966$~\AA). Arrows representing local dipoles are colored based on $P_x$ components in (\textbf{f}). The 90$^\circ$ domain walls separating $-P_y$ and $+P_z$ domains exhibit substantial $P_x$ components and adopt antiferroelectric coupling between neighboring walls. \textbf{g,} Spontaneous stochastic oscillating 90$^\circ$ domain walls in the absence of external electric fields. \textbf{h-i,} Dipole orientation distributions in strongly stretched $c/a$ domain structures.
 \label{ca2domain}}
\end{figure}

To comprehend the strain-dependent $d_{33}$ of the $c/a$ domain structure, we calculate the distributions of dipole (unit-cell polarization) orientations in both single $c$-domain and $c/a$ two-domain states at the same strain of $a_{\rm IP}=3.944$~\AA, using configurations sampled from equilibrium MD trajectories of at least 20 ps. This approach provides a statistical perspective on the dynamic structure.
The dipole orientation is gauged by its azimuthal angle ($\phi$) in the [111] plane (Fig.~\ref{ca2domain}\textbf{a}) to better distinguish $c$ ([001]) and $a$ ($[0\bar{1}0]$) domains. As shown in Fig.~\ref{ca2domain}\textbf{b-c}, the single $c$-domain features a $\phi$ distribution peaking at 45$^\circ$. In contrast, the dynamic structure of the $c/a$ state (Fig.~\ref{ca2domain}\textbf{d-e}) has a $\phi$ distribution ranging continuously from 0$^\circ$ and 360$^\circ$, with broadened peaks at 45$^\circ$ and 225$^\circ$, corresponding to $[001]$ and $[0\bar{1}0]$ dipoles, respectively. Dipoles with angle values deviating from the two main peaks are mainly near 90$^{\circ}$ domain walls, serving as continuously varying intermediate states bridging $a$ and $c$ domains. This marked difference in dynamic structure between the single $c$-domain and $c/a$ two-domain states is also evident in polar coordinates (Fig.~\ref{ca2domain}\textbf{b} and \textbf{d}).
In response to $\mathcal{E}_3$, dipoles in the single $c$-domain rotate away from the [001] direction, reducing the peak height at 45$^\circ$ (Fig.~\ref{ca2domain}\textbf{c}). In comparison, the same $\mathcal{E}_3$ induces more pronounced changes to the distribution in the $c/a$ state (Fig.~\ref{ca2domain}\textbf{e}), indicating that the enhanced $d_{33}$ results from the collective and coordinated small-angle rotations of dipoles at domain walls, analogous to ``coordinated gear dynamics", rather than the conventional understanding of 90$^\circ$ polarization rotation between domains~\cite{Noheda01p3891,Koo02p4205}. Additionally, the dipole orientation distribution associated with the dynamic structure does not show well-defined intermediate phases. 

The rapid rise in $d_{33}$ for the $c/a$ domain structure
beyond a critical tensile strain coincides with the emergence of substantial polarization components within domain walls~\cite{Wojdel14p247603}, \note{as well as a sharper increase in the domain wall thickness (see Fig.~S15)}. As illustrated in Fig.~\ref{ca2domain}\textbf{f}, domain walls separating $-P_y$ and $P_z$ domains exhibit $\pm P_x$  components, while adopting anti-parallel coupling between adjacent walls. 
Importantly, MD simulations reveal stochastic oscillations of these walls even without external driving forces (Fig.~\ref{ca2domain}\textbf{g}), suggesting minimal barriers for small-angle dipole rotations near domain walls. This is consistent with the diffuse distribution of dipole orientations in polar coordinates (Fig.~\ref{ca2domain}\textbf{h}) and the high susceptibility to $\mathcal{E}_3$ (Fig.~\ref{ca2domain}\textbf{i}). The mobile domain walls are responsible for the giant $d_{33}$ of $>250$ pC/N. \note{We note that the walls with $P_x$ components can be switched by an electric field applied along the $x$-axis, though the anti-parallel coupling between adjacent walls is favored thermodynamically (see Supplementary Sect.~V.E).}

\begin{figure}
	\begin{center}
		\includegraphics[width=0.8\textwidth]{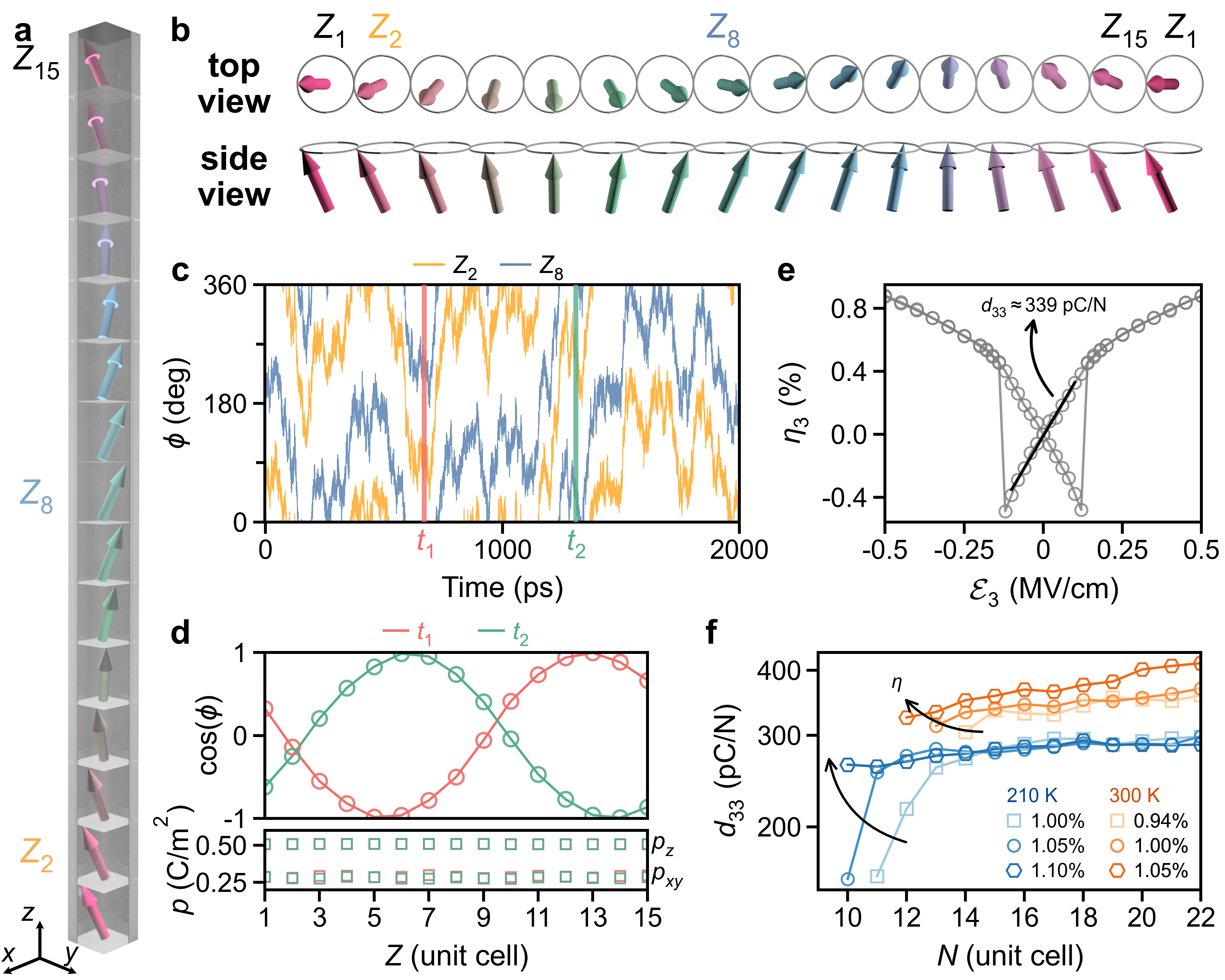}
	\end{center}
	\caption{\textbf{Helical dipole spiral in stretched PbTiO$_3$ membranes at 300~K.} \textbf{a-b,} Schematic illustrations of dipole ordering in the spiral. \textbf{c,} Evolution of instantaneous in-plane azimuth angles ($\phi$) of dipoles in two different $xy$ layers, $Z_2$ and $Z_8$, as labeled in (\textbf{a}). %\textbf{d,} Probability distribution of dipole orientation viewed along the out-of-plane direction. 
 \textbf{d,} Layer-resolved $\cos(\phi)$ and polarization profiles of instantaneous
configurations at $t_1$ and $t_2$ in (\textbf{c}). 
\textbf{e,} Strain-electric field ($\eta_3$-$\mathcal{E}_3$) hysteresis loops for dipole spirals.
\textbf{f,} \note{$d_{33}$ as a function of $N$ at varying strains and temperatures. The $y$-axis is in log scale for clarity.}
 \label{Spiral}}
\end{figure}

We now focus on the helical dipole spiral, which supports an even larger piezoelectric response ($d_{33}>$ 320 pC/N, see Fig.~\ref{MD}\textbf{c}). The non-collinear ordering of dipoles, obtained by averaging configurations over a 100-ps MD trajectory at 300~K for a 15$\times$15$\times$15 supercell, is depicted in Fig.~\ref{Spiral}\textbf{a}-\textbf{b}. The spiral, with a propagation vector aligned along [001] and a wavelength of $\approx$15 unit cells, is robust as confirmed by MD simulations using various supercell sizes (Fig.~S4).
Specifically, the dipoles, tilted by $\approx$27$^\circ$ from the $z$ axis (Fig.~\ref{Spiral}\textbf{a}), exhibit in-plane components that align collinearlly but rotate 24$^\circ$ relative to the preceding layer (Fig.~\ref{Spiral}\textbf{b}, top view); the out-of-plane components remain largely unchanged (Fig.~\ref{Spiral}\textbf{b}, side view). 

We find that the dynamic structure of the dipole spiral is quite vibrant in two aspects. Figure~\ref{Spiral}\textbf{c} tracks the evolution of the instantaneous in-plane azimuth angles ($\phi$) of dipoles in two different $xy$ layers (denoted as $Z_2$ and $Z_8$, respectively), 6 unit cells apart along the [001] direction. 
The $\phi$ value for each individual layer fluctuates stochastically due to thermal activation, but the angle difference consistently remains around 144$^\circ$, matching well with the expected 24$^\circ$ rotational difference per layer. 
Layer-resolved $\cos(\phi)$ and polarization profiles of instantaneous configurations at two different time points ($t_1$ and $t_2$, separated by 640~ps) are plotted in Fig.~\ref{Spiral}\textbf{d}, revealing the maintained helical configuration with shifted $\cos(\phi)$ profiles and unchanged polarization magnitudes.
These results indicate that the dipoles rotate collectively, coherently, and stochastically around the [001] direction and their collective response to external stimuli, achieved via small-angle rotations, is responsible for the giant piezoelectric effect. The simulated $\mathcal{\eta}_3$-$\mathcal{E}$ hysteresis loop, shown in Fig.~\ref{Spiral}\textbf{e}, further confirms the switchability of the dipole spiral (see MD snapshots in Fig.~S6) as well as the reversible electromechanical coupling. 
\note{This is distinct from the helical texture of electric dipoles in  BiCu$_{0.1}$Mn$_{6.9}$O$_{12}$, which exhibits almost no out-of-plane polarization ($<$ 20 $\mu$C/m$^2$) due to its improper nature~\cite{Khalyavin20p680}.}

\note{
We further investigate the effects of strain on the wavelength (measured in $N$ unit cells) of dipole spirals and the magnitude of $d_{33}$ at two different temperatures, 210 and 300~K (see MD versus experimental temperatures in Fig.~S13). 
%It is noted that a simulated temperature of 210~K in MD simulations effectively models the temperature of 300~K in experiments for PbTiO$_3$, given that the model potential underestimates the ferroelectric-paraelectric phase transition temperature. 
As shown in Fig.~\ref{Spiral}\textbf{f}, at 210~K and a tensile strain of 1\%, the dipole spiral has a minimum wavelength limit: spirals with $N<11$ will spontaneously transform into other domain structures in MD simulations, due to the increased gradient energy when $N$ becomes small.
Interestingly, dipole spirals with $N$ up to 22 are all stable, showing no spontaneous transformation during the equilibrium process. 
This stability aligns with predictions from a Landau-Ginzburg-Devonshire (LGD) model developed for the dipole spiral (see Supplementary Sect.~IV), which reveals a slow increase in free energy as $N$ increases. 
%Limited by the computational cost, we did not explore the stability of the dipole spiral for $N>22$ in MD simulations.
A larger tensile strain, such as 1.05\% and 1.10\%, reduces the minimum stable wavelength to $N=10$. 
Overall, the strain has a weak impact on the magnitude of $d_{33}$, which stabilizes at $\approx255$ pC/N at 210~K. Increasing the temperature to 300~K pushes the minimum stable wavelength to larger values. For example, at a tensile state of 1\%, we can only obtain dipole spirals with $N\ge 13$. The magnitude of $d_{33}$ becomes more sensitive to both strain and $N$ at 300~K, potentially achieving values greater than 400 pC/N. These results reveal a complex interplay between temperature, strain, wavelength, and piezoelectric response, highlighting the susceptible nature of dipole spirals.

Finally, we propose a feasible experimental approach to realize the dipole spiral. Our MD simulations of free-standing membranes of PbTiO$_3$, conducted with three-dimensional periodic boundary conditions, indicate that eliminating the depolarization field could facilitate the emergence of a dipole spiral. We design all-ferroelectric superlattices composed of alternating layers of PbTiO$_3$ and Pb$_{0.5}$Sr$_{0.5}$TiO$_3$ and find that this layered heterostructure supports arrays of dipole spirals in Pb$_{0.5}$Sr$_{0.5}$TiO$_3$ layers, each linking a pair of polar vortices within PbTiO$_3$ layers (see Fig.~S16). }

In summary, our findings demonstrate that dynamic structure dictates functional properties. For the extensively studied $c/a$ two-domain state in PbTiO$_3$, we suggest that the enhanced piezoelectric effect arises from the collective, small-angle dipole rotations near domain walls. A dipole spiral in tensile-strained PbTiO$_3$ membranes is discovered, representing a new state of polar ordering with strongly correlated dipoles that can rotate freely without energy cost, indicative of a zero-energy mode. This topological polar structure offers an avenue for enhancing electromechanical coupling and exploring phenomena such as chiral phonon dynamics~\cite{Zhu18p579} and non-collinear ferroelectricity~\cite{Zhao21p341}.

%\begin{acknowledgments}
%Y.H., J.Y., and S.L. acknowledge the supports from Natural Science Foundation of Zhejiang Province (2022XHSJJ006) and Westlake Education Foundation. We acknowledge useful discussions with Prof. Hongjian Zhao. The computational resource is provided by Westlake HPC Center. 
%\end{acknowledgments}

\bibliography{SL.bib}

\clearpage
\newpage

 \clearpage
\newpage

 \clearpage
\newpage

 \clearpage
\newpage

\end{document}

\title{Supplemental Material for \\ Giant piezoelectric effects of topological structures in stretched ferroelectric membranes}
% \keywords{Keywords: ...}
\author{Yihao Hu}
\affiliation{Key Laboratory for Quantum Materials of Zhejiang Province, Department of Physics, School of Science, Westlake University, Hangzhou, Zhejiang 310024, China}
\author{Jiyuan Yang}
\affiliation{Key Laboratory for Quantum Materials of Zhejiang Province, Department of Physics, School of Science, Westlake University, Hangzhou, Zhejiang 310024, China}
\author{Shi Liu}
\email{liushi@westlake.edu.cn}
\affiliation{Key Laboratory for Quantum Materials of Zhejiang Province, Department of Physics, School of Science, Westlake University, Hangzhou, Zhejiang 310024, China}
\affiliation{Institute of Natural Sciences, Westlake Institute for Advanced Study, Hangzhou, Zhejiang 310024, China}
\maketitle

\clearpage

\linespread{1.36}

\section{Computational Methods}
\subsection{DEEP POTENTIAL FROM DFT}
The deep potential (DP) is a deep neural network-based model potential that maps the local environment of atom $i$ to its atomic energy ($E_i$). The total energy is the sum of these atomic energies, $E =\sum _i E_i$. The DP model used in this work is trained on a database of DFT energies and atomic forces for 19119 Pb$_{x}$Sr$_{1-x}$TiO$_3$ configurations constructed using 40-atom  $2\times2\times2$ supercells. The final training database contains three datasets:
\begin{itemize}
    \item PbTiO$_3$: the converged PbTiO$_3$ database consists of 13,021 configurations including 40-atom $2\times 2\times2$ supercells of tetragonal $P4mm$ and cubic ($Pm\bar{3}m$) phases. 
    \item SrTiO$_3$: we use a published database with 3,538 configurations including 40-atom $Pm\bar{3}m$ supercells and 20-atom $I4/mcm$ supercells~\cite{He22p064104}.
    \item Pb$_x$Sr$_{1-x}$TiO$_3$ solid solutions: this dataset is generated via a concurrent learning procedure and includes 2,560 configurations of Pb$_x$Sr$_{1-x}$TiO$_3$ with $x=0.25$, 0.50, and 0.75.
\end{itemize}

All DFT calculations are carried out using the Vienna Ab initio Simulation (VASP) package with the projected augmented wave method and the Perdew-Burke-Ernzerhof functional modified for solids (PBEsol). 
An energy cutoff of 800 eV and a $k$-spacing is 0.3~\AA$^{-1}$ are enough to converge the energy and atomic forces. 
Additional details, including the construction of the database, training protocol, and metadata of the model, were documented in the previous work~\cite{Wu23p144102}.
The DP model of Pb$_{x}$Sr$_{1-x}$TiO$_3$ is capable of predicting various properties of solid solutions, such as the phonon spectra of different phases of PbTiO$_3$ and SrTiO$_3$, temperature-driven and composition-driven phase transitions. 
In particular, it reproduces an in-plane strain-driven transition from an ordered polar vortex lattice to a shifted polar vortex lattice, and to electric dipole waves in PbTiO$_3$/SrTiO$_3$ superlattices, highlighting its transferability to model different strain states and complex domain structures~\cite{Wu23p144102}.

\begin{figure}[htbp]
\includegraphics[width=0.95\textwidth]{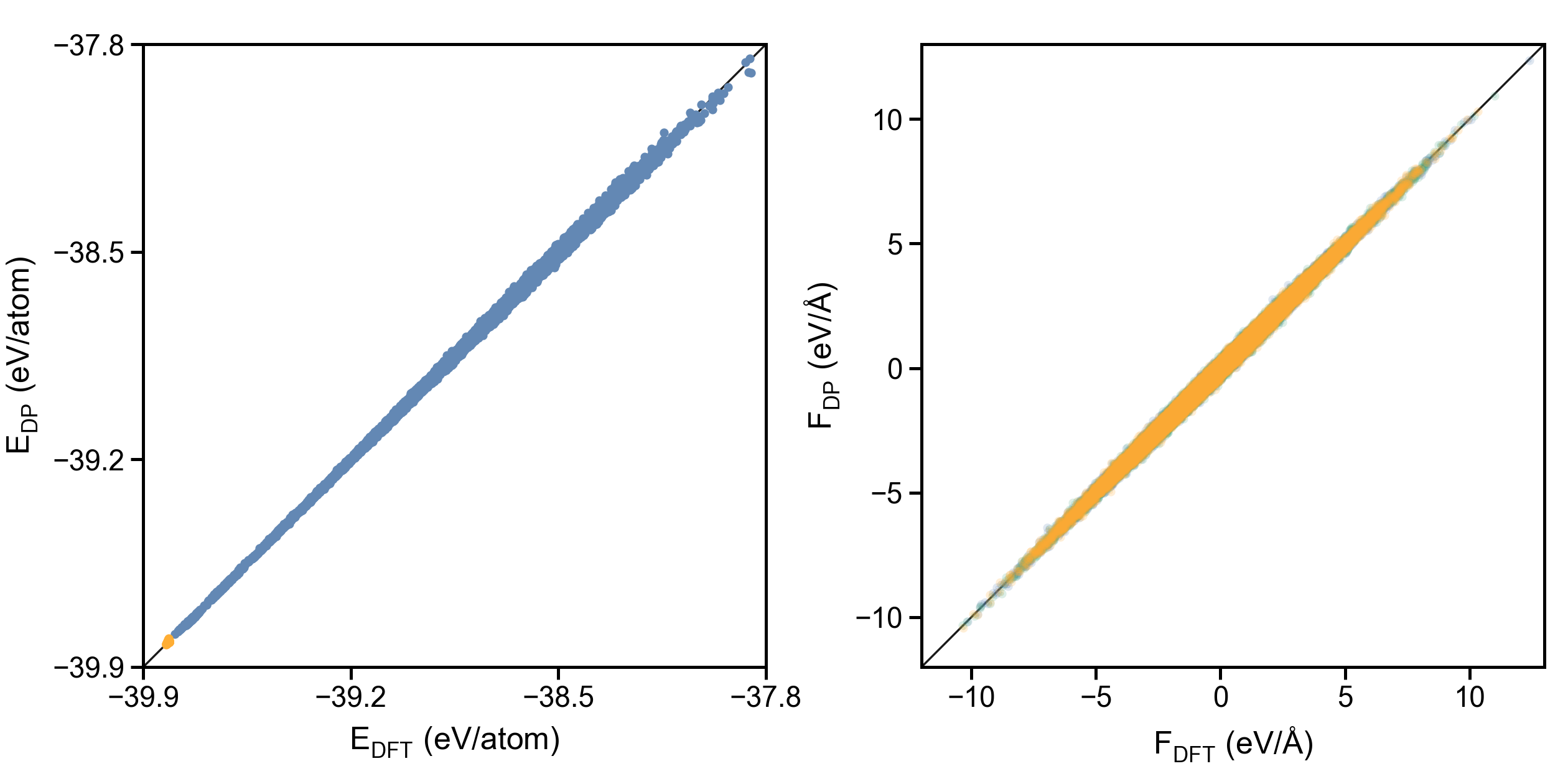}
\caption{Comparison of (a) energies and (b) atomic forces computed with DFT and DP for
all configurations in the training database. The DP model accurately predicts the energies of configurations of 5-atom unit cells (yellow points in (a)) that are not included in the training database. These configurations were used to construct the multiphase diagram presented in Figure 1 of the main text.
}
\label{fig_phases}
\end{figure}

Figure~\ref{fig_phases} presents a comparison between the energies and atomic forces as calculated by both DP and DFT for all configurations in the training database, demonstrating the DP model's excellent fit to the DFT results.
We note that the DP model accurately reproduces the energies of all configurations involved in the multiphase diagram (Figure 1 in the main text), which are not in the training database. We have developed an online notebook on Github (\href{https://github.com/huiihao/Spiral}{https://github.com/huiihao/Spiral}) that publishes the training database, force field model, training metadata, essential input files for DFT calculations and MD simulations, data analysis scripts, and selected original MD trajectories. 
%\clearpage

\tikzset{every picture/.style={line width=1.75pt}} %设置默认线宽为1.75pt        
\begin{tikzpicture}[remember picture, overlay]
\node at (current page.west) [anchor=east, xshift=-4.5cm, yshift=11.8cm] {
\begin{tikzpicture}[remember picture, overlay, x=1.05pt,y=1.05pt,yscale=-1,xscale=1]
\node [anchor=base, inner sep=0pt, font=\bfseries\fontsize{14}{16}\selectfont\fontfamily{arial}] at (208,105) {a};
\node [anchor=base, inner sep=0pt, font=\bfseries\fontsize{14}{16}\selectfont\fontfamily{arial}] at (424,105) {b};
\end{tikzpicture}
};
\end{tikzpicture}

\clearpage
\newpage

\subsection{Molecular dynamics simulations}
The misfit strain-temperature domain stability diagram is constructed by running DPMD simulations in the isobaric-isothermal ($NPT$) ensemble with in-plane lattice constants fixed. All MD simulations are performed using \texttt{LAMMPS} \cite{Plimpton95p1}, with temperature controlled via the Nos\'e-Hoover thermostat and pressure controlled by the Parrinello-Rahman barostat. The timestep for the integration of the equation of motion is 2 fs. The pressure is maintained at 1.0 bar along the out-of-plane direction and the temperature ranges from 210~K to 420~K. At a given temperature, the equilibrium run lasts more than 50 ps, followed by a production run of 50 ps that is sufficiently long to obtain converged statistical descriptions of dynamic structures.

The single-domain state and the dipole spiral are modeled using 
15$\times$15$\times$15 perovskite-type supercells containing 16,875 atoms.
Larger systems, such as 15$\times$15$\times$25 and 25$\times$25$\times$15 supercells, are used to verify the robustness of the dipole spiral at 300~K.
The 4$\times$40$\times$40 and 40$\times$40$\times$4 supercells, each comprising 32,000 atoms, are adopted to model the $c/a$ and $a_1/a_2$ two-domain states, respectively, ensuring that the domain size is consistent in these two domain structures. In the calculations of piezoelectric coefficient $d_{33}$, electric fields are included in classical MD simulations using the ``force method", where an additional force $\mathcal{F}_i$ is applied to ion $i$ according to $\mathcal{F}_i = Z_i^*\cdot \mathcal{E}$, with $Z_i^*$ representing the Born effective charge tensor of ion $i$ computed with DFT. The polarization of the unit cell is estimated using the following formula,
\[\mathbf{p}^m(t)=\frac{1}{V_{\rm uc}}\left[\frac{1}{8} \mathbf{Z}_{\mathrm{Pb}}^* \sum_{k=1}^8 \mathbf{r}_{\mathrm{Pb}, k}^m(t)+\mathbf{Z}_{\mathrm{Ti}}^* \mathbf{r}_{\mathrm{Ti}}^m(t)+\frac{1}{2} \mathbf{Z}_{\mathrm{O}}^* \sum_{k=1}^6 \mathbf{r}_{\mathrm{O}, k}^m(t)\right]\]
where $\mathbf{p}^m(t)$ is the polarization of unit cell $m$ at time $t$, $V_{\rm uc}$ is the volume of the unit cell, $\mathbf{Z}_{\mathrm{Pb}}^*, \mathbf{Z}_{\mathrm{Ti}}^*$, and $\mathbf{Z}_{\mathrm{O}}^*$ are the Born effective charges of Pb, Ti and O atoms, $\mathbf{r}_{\mathrm{Pb}, k}^m(t), \mathbf{r}_{\mathrm{Ti}}^m(t)$, and $\mathbf{r}_{\mathrm{O}, k}^m(t)$ are the instantaneous atomic positions in unit cell $m$ from MD simulations. Here, the local polarization $\mathbf{p}^m$ is defined as the local electric dipole divided by $V_{\rm uc}$.

\clearpage
\newpage

\section{Three-dimensional real-space dipole distributions of dipole spiral}

\subsection{Dipole spiral modeled with a  $15\times15\times15$ supercell}
\begin{figure}[htbp]
  % \centering
  \includegraphics[width=0.98\textwidth]{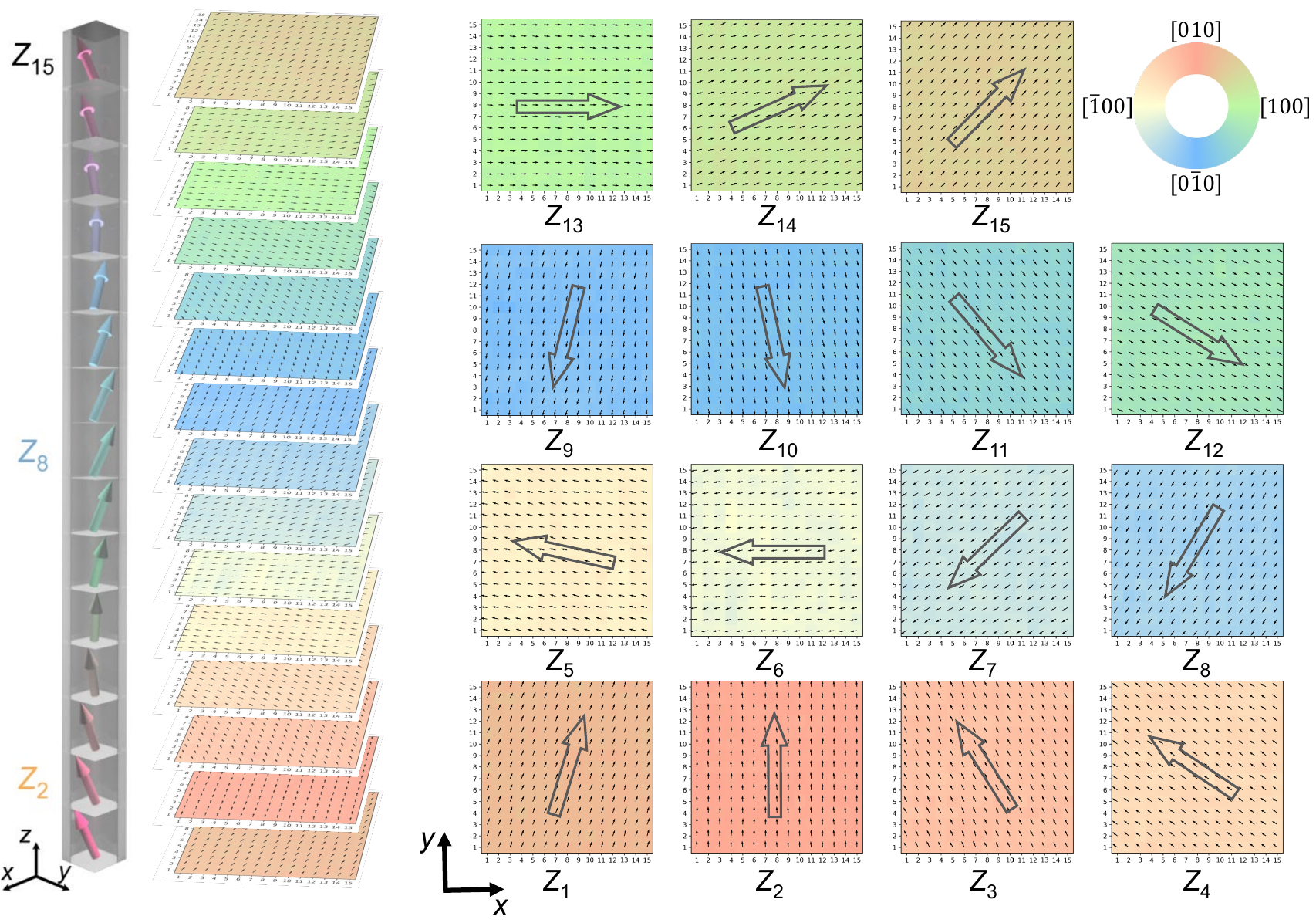}
  \caption{Layer-resolved dipole distributions of a dipole spiral. The structure is analyzed by dissecting the dipoles across all $xy$ planes. Each black arrow represents the in-plane components of local electric dipole within a unit cell, with the background color illustrating the dipole direction.}
  \label{xySpiral}
\end{figure}

 We present in Fig.~\ref{xySpiral} the layer-resolved dipole distributions of a typical dipole spiral, which is modeled using a $15\times15\times15$ supercell at 300 K with in-plane unit-cell lattice constants fixed at 3.958~\AA. 
The spiral propagates along the [001] direction (the $z$ axis) and features a wavelength of $\approx$15 unit cells. The dipoles are tilted at an angle of $\approx$27$^\circ$ from the $z$ axis, and their in-plane components on each $xy$ plane align collinearly.

\begin{figure}[htbp]
\includegraphics[width=0.98\textwidth]{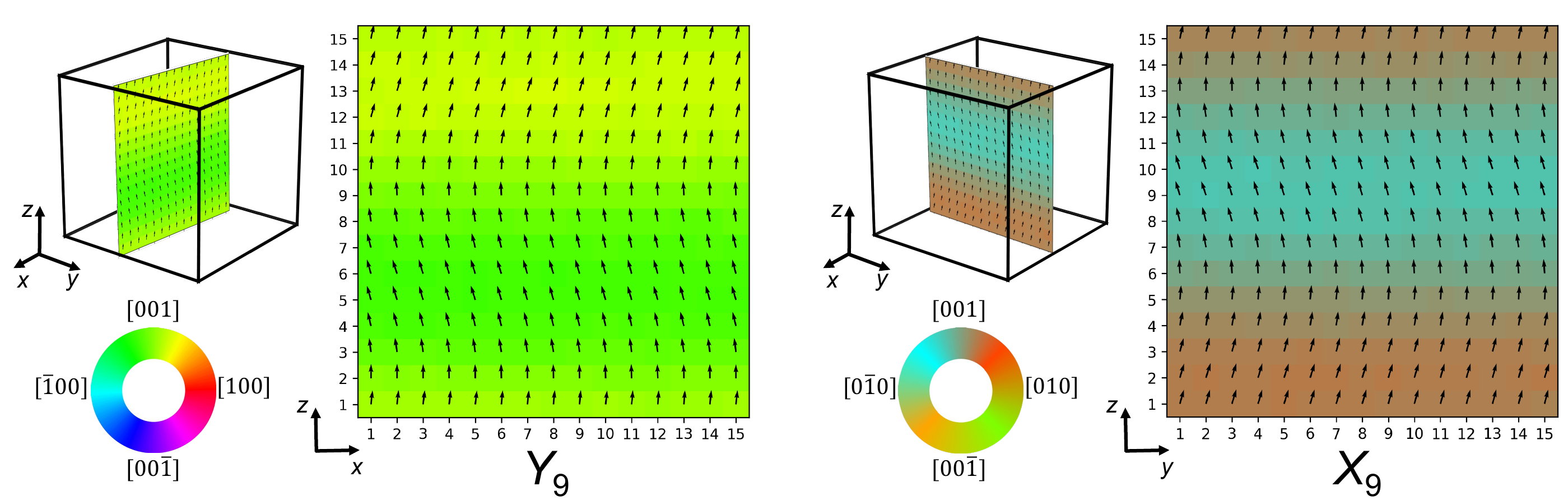}
  \caption{Dipole-wave patterns in representative (a) $xz$ cross-section and (b) $yz$ cross-section of 3D dipole distributions of a dipole spiral.}
  \label{crossSprial}
\end{figure}

\clearpage
\newpage
Figure~\ref{crossSprial} presents specific $xz$ and $yz$ cross-sections of the dipole distributions, denoted as $Y_9$ and $X_9$, respectively, revealing periodic electric dipole waves characterized by head-to-tail connected electric dipoles in the form of a sinusoidal function. 
The dipole-wave patterns in $xz$ and $yz$ cross-sections represent the projected views of a three-dimensional (3D) helical dipole spiral. These patterns appear similar across all $xz$ and $yz$ planes.
Practically, the presence of a dipole spiral can be conveniently ascertained by analyzing the dipole pattern in either the $xz$ or $yz$ cross-sections.
It is evident that the out-of-plane component along the $z$ axis remains nearly unchanged.

\tikzset{every picture/.style={line width=1.75pt}} %设置默认线宽为1.75pt        
\begin{tikzpicture}[remember picture, overlay]
\node at (current page.west) [anchor=east, xshift=-4.5cm, yshift=11.8cm] {
\begin{tikzpicture}[remember picture, overlay, x=1.05pt,y=1.05pt,yscale=-1,xscale=1]
\node [anchor=base, inner sep=0pt, font=\bfseries\fontsize{14}{16}\selectfont\fontfamily{arial}] at (207,183) {a};
\node [anchor=base, inner sep=0pt, font=\bfseries\fontsize{14}{16}\selectfont\fontfamily{arial}] at (431,183) {b};
\end{tikzpicture}
};
\end{tikzpicture}

\clearpage
\newpage
\subsection{Supercell size effect on the formation of dipole spiral}
We perform MD simulations using supercells of various sizes to examine the robustness of the emergence of the dipole spiral. All these simulations are performed at 300~K with in-plane unit-cell lattice constants fixed at 3.958~\AA. The real-space dipole distributions are calculated for the ensemble-averaged structure, derived by averaging configurations over a 100-ps MD trajectory.
Figure~S4 plots the $xz$ cross-sections of dipole distributions for various supercells, consistently demonstrating dipole waves that are indicative of helical dipole spirals.

\begin{figure}[htbp]
\includegraphics[width=0.98\textwidth]{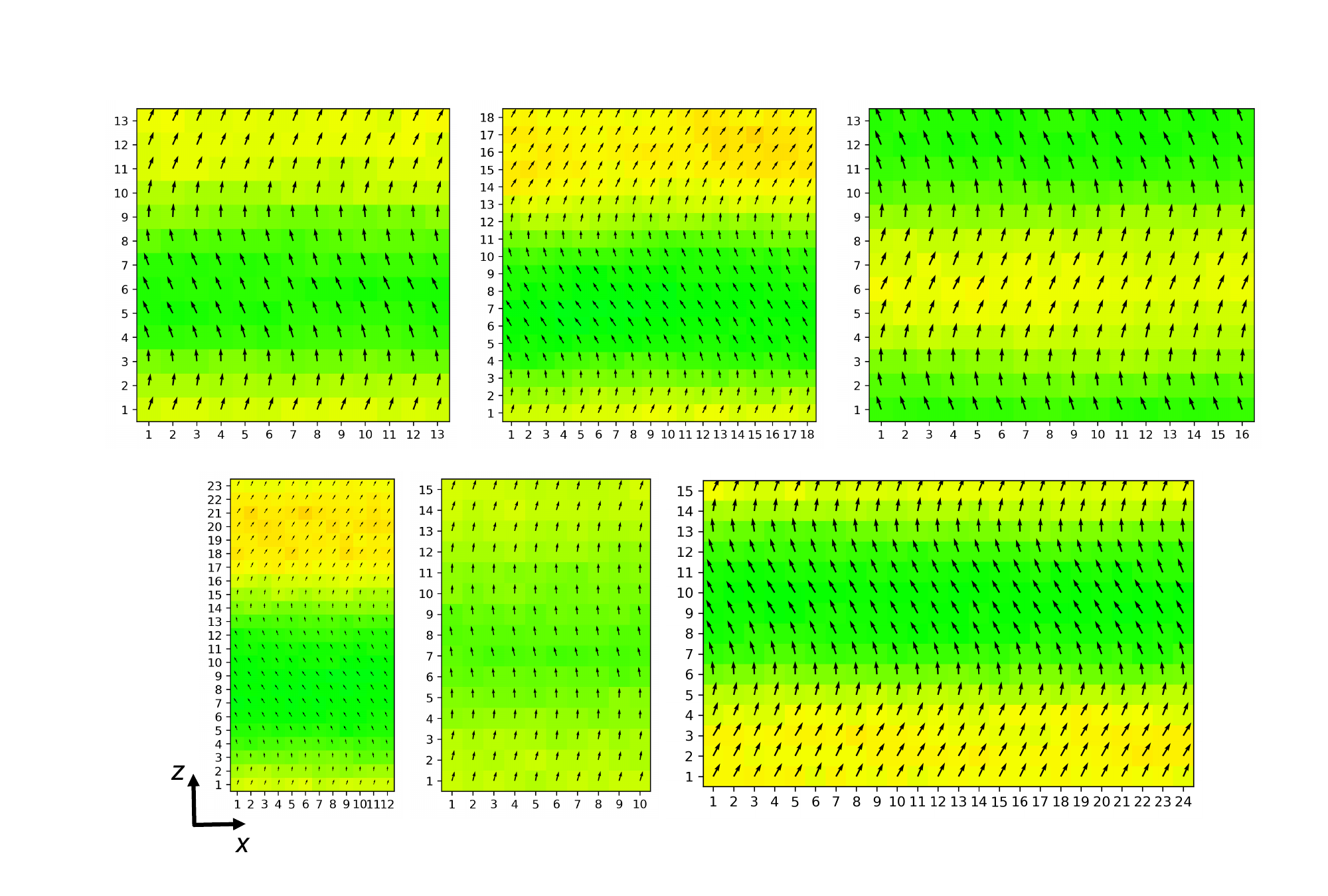}
   \caption{Representative $xz$ cross-sections of 3D dipole distributions of dipole spirals modeled with (a) $13\times13\times13$, (b) $18\times18\times18$, (c) $16\times16\times13$, (d) $12\times12\times23$, (e) $10\times10\times15$, and (f) $24\times24\times15$ supercells.}
\end{figure}

\tikzset{every picture/.style={line width=1.75pt}} %设置默认线宽为1.75pt        
\begin{tikzpicture}[remember picture, overlay]
\node at (current page.west) [anchor=east, xshift=-4.5cm, yshift=11.8cm] {
\begin{tikzpicture}[remember picture, overlay, x=1.05pt,y=1.05pt,yscale=-1,xscale=1]
\node [anchor=base, inner sep=0pt, font=\bfseries\fontsize{14}{16}\selectfont\fontfamily{arial}] at (206,198) {a};
\node [anchor=base, inner sep=0pt, font=\bfseries\fontsize{14}{16}\selectfont\fontfamily{arial}] at (340,198) {b};
\node [anchor=base, inner sep=0pt, font=\bfseries\fontsize{14}{16}\selectfont\fontfamily{arial}] at (477,198) {c};
\node [anchor=base, inner sep=0pt, font=\bfseries\fontsize{14}{16}\selectfont\fontfamily{arial}] at (240,335) {d};
\node [anchor=base, inner sep=0pt, font=\bfseries\fontsize{14}{16}\selectfont\fontfamily{arial}] at (318,335) {e};
\node [anchor=base, inner sep=0pt, font=\bfseries\fontsize{14}{16}\selectfont\fontfamily{arial}] at (414,335) {f};
\end{tikzpicture}
};
\end{tikzpicture}

\clearpage
\newpage
\section{Piezoelectric response of dipole spiral}
\subsection{Computing $d_{33}$ with finite-field MD simulations}
\begin{figure}[htbp]
\includegraphics[width=0.6\textwidth]{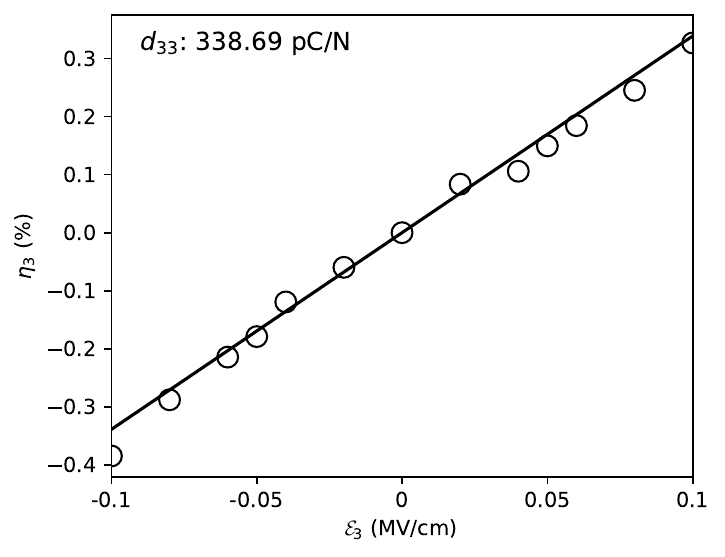}
\caption{Strain change ($\eta_3$) of a dipole spiral in response to an out-of-plane electric field ($\mathcal{E}_3$). All MD simulations are performed at 300~K with in-plane unit-cell lattice constants fixed at 3.958~\AA.}
\label{Spiral-d33}
\end{figure}

The piezoelectric coefficient \(d_{33}\) of a dipole spiral is estimated based on the direct piezoelectric effect, $[\partial{\eta}_3/\partial{\mathcal{E}_3}]|_{\sigma_3=0}$, where \(\eta_3\) denotes the strain change due to an out-of-plane electric field (\(\mathcal{E}_3\)). As shown in Fig.~\ref{Spiral-d33}, the strain exhibits a linear dependence when \(\mathcal{E}_3\) is within the range of \(-0.1\) MV/cm to \(0.1\) MV/cm. The value of \(d_{33}\), computed from the slope of \(\eta_3\) versus \(\mathcal{E}_3\), is \(339\) pC/N.

\clearpage
\newpage
\subsection{Strain-electric field hysteresis loop}
The dipole spiral is robust against the application of an external electric field. We observe a reversible $\mathcal{E}_3$-driven transition from a spiral to a single $c$-domain. Specifically, when the magnitude of $\mathcal{E}_3$ is above 0.2 MV/cm, the helical spiral evolves into a single-domain state with polarization aligned along the $z$ axis. Upon removal of the electric field, the $c$ domain spontaneously evolves to a dipole spiral. The hysteresis loops also confirm the switchability of the dipole spiral (see snapshots 1 and 5).

\begin{figure}[htbp]
\includegraphics[width=0.98\textwidth]{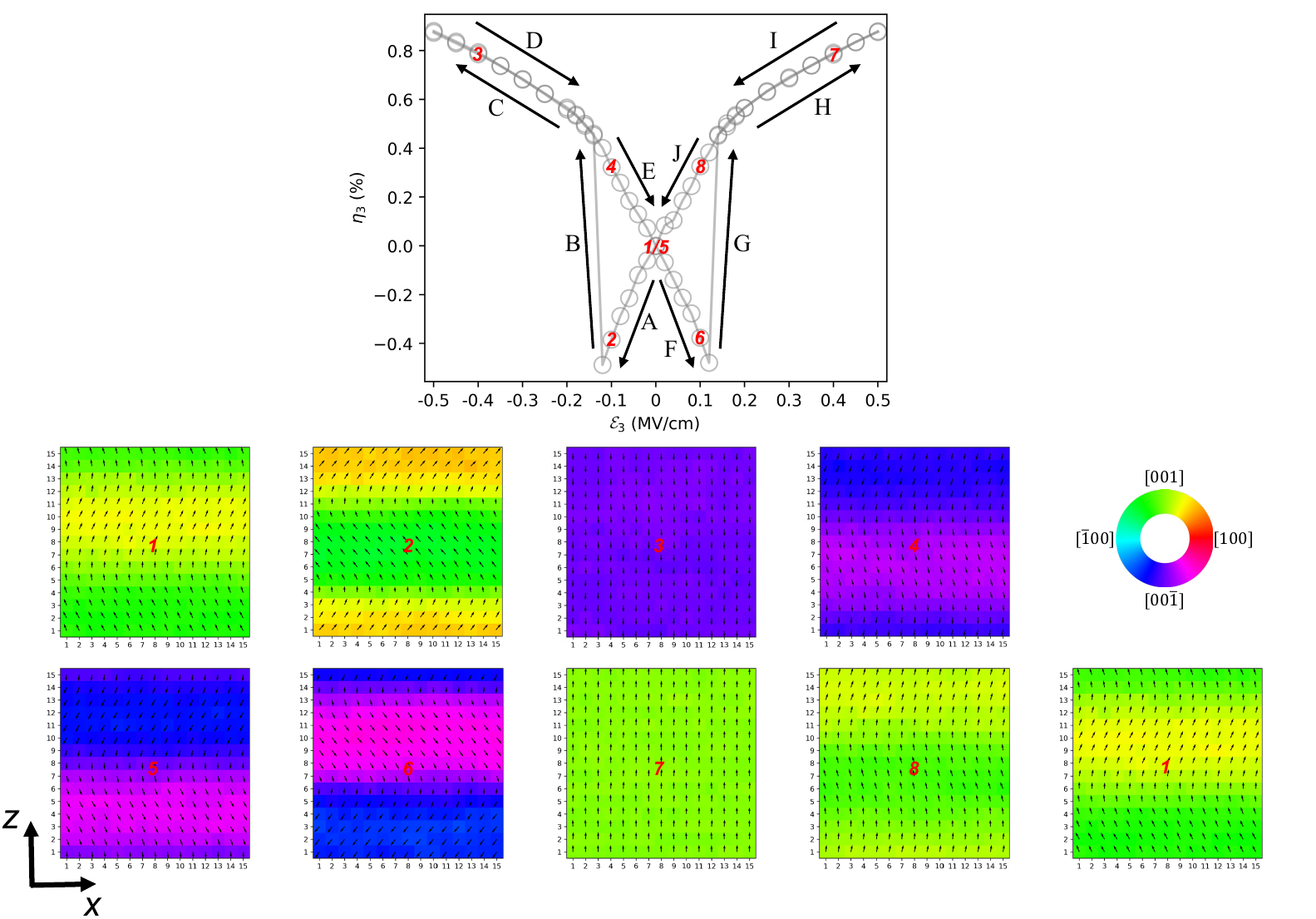}
  \caption{Strain-electric field ($\eta_3$-$\mathcal{E}_3$) hysteresis loops for dipole spirals simulated with DPMD at 300~K. The $xz$ cross-sections of 3D dipole distributions for various states along the loop reveal reversible transitions between the dipole spiral and the single $c$-domain state.}
\end{figure}

\clearpage
\newpage

\section{Landau-Ginzburg-Devonshire model}

\subsection{Total free energy}
The three-dimensional polarization distribution of a dipole spiral propagating along the $z$ axis can be described with the following parameters: in-plane polarization $p_{xy}=(p_x,p_y)$ in each $xy$ plane, out-of-plane polarization $p_z$, wavelength $N$ in unit cells, in-plane azimuth angle $\phi_0$ of $p_{xy}$ in an arbitrary starting $xy$ layer (denoted as $Z_1$). 
The direction of $p_{xy}$ rotates by $\delta=2\pi/N$ progressively relative to the preceding $xy$ layer, and the magnitudes of $p_{xy}$ and $p_z$ are the same across all layers. 
The total free energy ($F$) of a dipole spiral modeled with a $1\times1\times N$ supercell is given by:
\begin{equation}
    F=\sum_{k=1}^N f_{\rm loc}^k + \sum_{k=1}^N f_g^k
\end{equation}
where $f_{\rm loc}^k$ is the local energy contribution of layer $i$ from the Landau–Devonshire phenomenological theory, and $f_g^k$ is the gradient energy due to the polarization discontinuity between $xy$ layers $k$ and $k+1$. Following the treatment of PbZr$_{x}$Ti$_{1-x}$O$_3$, we express $f_{\rm loc}^k$ as a sixth-order polynomial:
\begin{align}
    f_{\rm loc}^k(p_x, p_y, p_z) = &\alpha_1(p_x^2+p_y^2+p_z^2)+\alpha_{11}(p_x^4+p_y^4+p_z^4)+\alpha_{12}(p_x^2p_y^2+p_y^2p_z^2+p_x^2p_z^2) \nonumber \\
    &+\alpha_{111}(p_x^6+p_y^6+p_z^6)+\alpha_{112}[p_x^2(p_y^4+p_z^4)+p_y^2(p_x^4+p_z^4)+p_z^2(p_x^4+p_y^4)] \nonumber  \\
    &+\alpha_{123}p_x^2p_y^2p_z^2 \label{sixth}
\end{align}
where $\alpha_1$, $\alpha_{11}$, $\alpha_{12}$, $\alpha_{111}$, $\alpha_{112}$, and $\alpha_{113}$ are Landau–Devonshire coefficients. Using $p_x=p_{xy}\cos\phi_k$ and $p_y=p_{xy}\sin\phi_k$, equation~(\ref{sixth}) becomes:
\begin{equation}
\begin{aligned}
    f_{\rm loc}^k(p_{xy}, p_z, \phi_k) =
    &\alpha_1p_{xy}^2+\alpha_1 p_z^2 +\alpha_{11}p_z^4+\alpha_{12}p_{xy}^2p^2_z+\alpha_{111}p_z^6+\alpha_{112}p_{xy}^2p_z^4 \\
    & +p_{xy}^4\frac{1}{8}[(2\alpha_{11}-\alpha_{12})\cos(4\phi_k)+(6\alpha_{11}+\alpha_{12})] \\
    & +p_{xy}^6\frac{1}{8}[(3\alpha_{111}-\alpha_{112})\cos(4\phi_k)+(5\alpha_{111}+\alpha_{112})]\\
    & +p_{xy}^4p_z^2\frac{1}{8}[(2\alpha_{112}-\alpha_{123})\cos(4\phi_k)+(6\alpha_{112}+\alpha_{123})]\\
\end{aligned}
\label{floc2}
\end{equation}
where $\phi_k=\phi_0+k\frac{2\pi}{N}$.  We note that the choice of a sixth-order polynomial for the local energy is based on the probability distribution of in-plane dipole components of the dynamic structure (Fig.~S7), which indicates a quadruple-well energy landscape within the $xy$ plane.

The gradient energy is:
\begin{equation}
    f_g^k = g_{x}[p_{xy}\cos\phi_k-p_{xy}\cos(\phi_k+\delta)]^2 + g_y[p_{xy}\sin\phi_k-p_{xy}\sin(\phi_k+\delta)]^2
\end{equation}
where $g_x$ and $g_y$ are gradient coefficients. 
Exploiting the in-plane isotropy, where $g_x = g_y = g$, we can simplify the expression for the gradient energy $f_g^k$. This yields $f_g^k = 4g p_{xy}^2 \sin^2(\delta/2) = 4g p_{xy}^2 \sin^2(\pi/N)$. 

Consequently, the total free energy is reformulated as a function of $p_{xy}$, $p_z$, $N$, and $\phi_0$:
\begin{equation}
    F(p_{xy}, p_z, N, \phi_0) = \sum_{k=1}^N f_{\mathrm{loc}}^k(p_{xy}, p_z, N, \phi_0) + \sum_{k=1}^N f_g^k(p_{xy}, N)
    \label{F2}
\end{equation}
In principle, with all Landau-Ginzburg-Devonshire coefficients known, one can search for the global free energy minima using equation~(\ref{F2}). 

\subsection{Simplified LGD model}

For simplicity, we assume $p_{xy}$ and $p_z$ already adopt their respective optimal values, $p_{xy}^s$ and $p_z^s$. The local free energy per layer, $f_{\rm loc}$, can be reformulated as:
\begin{equation}
\begin{aligned}
f_{\rm loc}=\frac{1}{N}\sum_i^Nf_{\rm loc}^k(\phi_0; p_{xy} = p_{xy}^s, p_z = p_z^s) = \frac{\mathcal{A}}{N} \sum_{k=1}^N \cos\left(4\left(\phi_0 + k\frac{2\pi}{N}\right)\right) + \mathcal{B}
\end{aligned}
\end{equation}
where
\begin{equation}
\begin{aligned}
\mathcal{A}=&~\frac{1}{8}p_{xy}^4(2\alpha_{11}-\alpha_{12})+ \frac{1}{8}p_{xy}^6(3\alpha_{111}-\alpha_{112})+ \frac{1}{8}p_{xy}^4 p_z^2(2\alpha_{112}-\alpha_{123})\\
\mathcal{B}=&~\alpha_1 p_{xy}^2+\alpha_1 p_z^2+\alpha_{12}p_{xy}^2p_z^2+\alpha_{11}p_z^4+\alpha_{111}p_z^6+\alpha_{112}p_{xy}^2p_z^4\\
&~+\frac{1}{8}p_{xy}^4(6\alpha_{11}+\alpha_{12})+\frac{1}{8}p_{xy}^6(5\alpha_{111}+\alpha_{112})+\frac{1}{8}p_{xy}^4 p_z^2(6\alpha_{112}+\alpha_{123})
\end{aligned}
\end{equation}
with $p_{xy}=p_{xy}^s$ and $p_z=p_z^s$. After performing the sum, we obtain: 
\begin{equation}
\begin{aligned}
   f_{\rm loc} &=
    \begin{cases}
       \mathcal{A}\cos(4\phi_0) +\mathcal{B}, & N=1,2,\text{or } 4,\\
       \mathcal{B}, & N=3,\text{or} >4.
    \end{cases}
    \label{Flocal4}
\end{aligned} 
\end{equation}

\begin{figure}[htbp]
\includegraphics[width=0.95\textwidth]{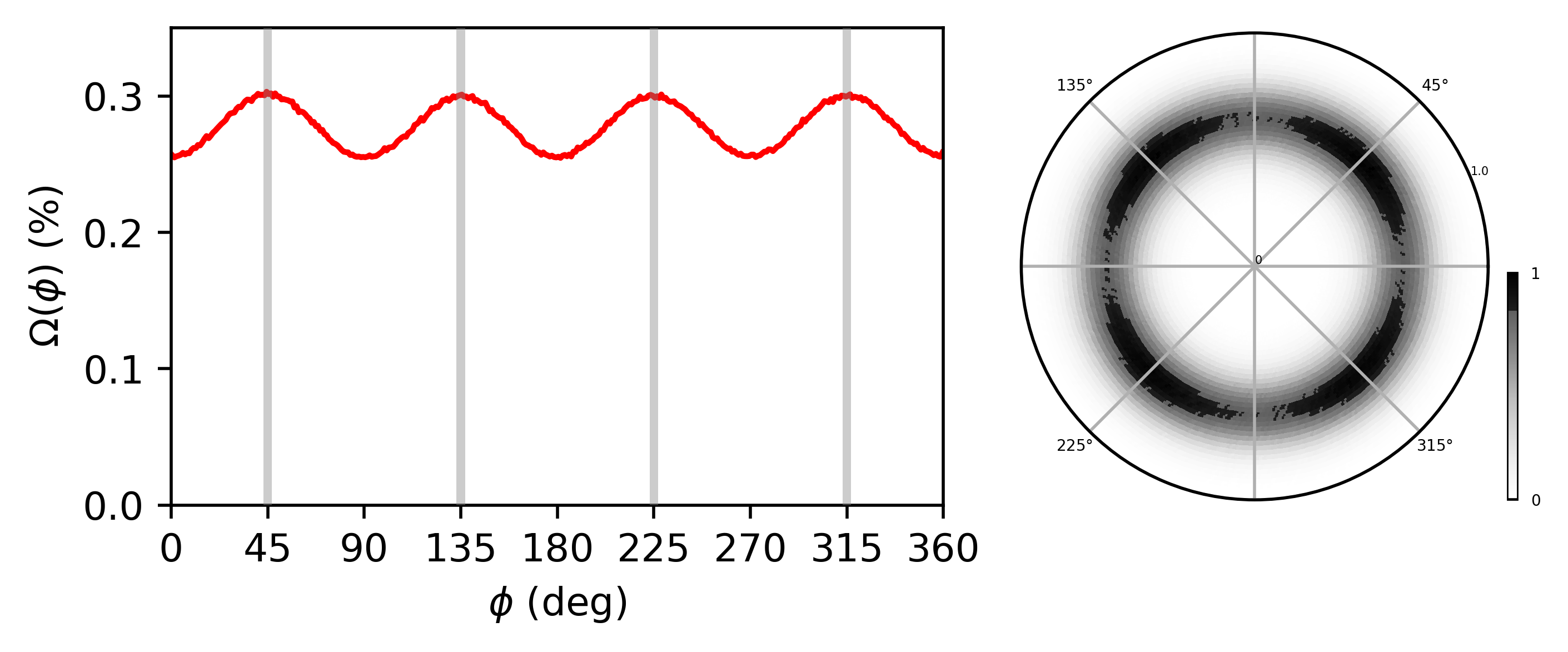}
\caption{Dipole orientation distributions of a spin spiral modeled with a  15$\times$15$\times$15 supercell. A 200 ps MD trajectory is used. (a) Distribution of in-plane azimuthal angles ($\phi$) in the \{001\} plane. (b) Dipole orientation distribution plotted in the polar coordinates viewed along [001].}
\label{quad}
\end{figure}

During the derivation,  the identity, $\sum_{k=1}^N \cos\left(4\left(\phi_0 + k\frac{2\pi}{N}\right)\right) = 0$, is used (see proof in APPENDIX).
In the case of $N=4$, it is easy to show that the local energy reaches the minimum value of $-\mathcal{A}+\mathcal{B}$ when $\phi_0$ adopts one of the four values, 45$^\circ$, 135$^\circ$, 225$^\circ$, 315$^\circ$. This is consistent with the in-plane quadruple-well energy landscape revealed from MD simulations at 300~K (Fig.~\ref{quad}). In the case of $N>4$, $f_{\rm loc}=\mathcal{B}$, which is independent of both the wavelength and $\phi_0$. 

\tikzset{every picture/.style={line width=1.75pt}} %设置默认线宽为1.75pt        
\begin{tikzpicture}[remember picture, overlay]
\node at (current page.west) [anchor=east, xshift=-4.5cm, yshift=11.8cm] {
\begin{tikzpicture}[remember picture, overlay, x=1.05pt,y=1.05pt,yscale=-1,xscale=1]
\node [anchor=base, inner sep=0pt, font=\bfseries\fontsize{14}{16}\selectfont\fontfamily{arial}] at (207,19) {a};
\node [anchor=base, inner sep=0pt, font=\bfseries\fontsize{14}{16}\selectfont\fontfamily{arial}] at (481,19) {b};
\end{tikzpicture}
};
\end{tikzpicture}

\iffalse
The total free energy then becomes:
\begin{equation}
    F(N, \phi_0; p_{xy} = p_{xy}^s, p_z = p_z^s) = \mathcal{A} \sum_{k=1}^N \cos\left(4\left(\phi_0 + k\frac{2\pi}{N}\right)\right) + \mathcal{B}N + \mathcal{C}N \sin^2\left(\frac{\pi}{N}\right)
    \label{F3}
\end{equation}
where $\mathcal{A}$, $\mathcal{B}$, and $\mathcal{C}$ are renormalized coefficients, which depend on the Landau-Ginzburg-Devonshire coefficients as well as on $p_{xy}^s$ and $p_z^s$ (see explicit expressions in Supplementary Section VI). The first two terms in equation~(\ref{F3}) correspond to the local energy, and the third term represents the gradient energy.
\fi

The total free energy per layer is given as
\begin{equation}
\begin{aligned}
    f&=f_{\rm loc} + f_g =\frac{1}{N}F(N, \phi_0;p_{xy}=p_{xy}^s, p_z=p_z^s )\\
    &= \left(\mathcal{A}\frac{1}{N} \sum_{k=1}^N\cos\left(4(\phi_0+k\frac{2\pi}{N})\right) + \mathcal{B}\right)+\mathcal{C}\sin^2\left(\frac{\pi}{N}\right)\\
    &=
    \begin{cases}
       \mathcal{A}\cos(4\phi_0) +\mathcal{B}+\mathcal{C}\sin^2\left(\frac{\pi}{N}\right), & N=1,2,\text{or } 4,\\
       \mathcal{B}+\mathcal{C}\sin^2\left(\frac{\pi}{N}\right), & N=3,\text{or} >4.
    \end{cases}
    \label{F4}
\end{aligned} 
\end{equation}
where $\mathcal{C}=4g(p_{xy}^s)^2$. The validity of equation~(\ref{F4}) is demonstrated by its excellent fitting to the DPMD energies of dipole spirals with various wavelengths ($4<N<55$) at 0~K, as shown in Fig.~\ref{Fitting}. 

%In the case where $N > 4$, the identity $\sum_{k=1}^N \cos\left(4\left(\phi_0 + k\frac{2\pi}{N}\right)\right) = 0$ is utilized (proof provided in Supplementary Section IX); consequently, the layer-averaged local energy contribution is $f_{\mathrm{loc}}=\mathcal{B}$, independent of both $N$ and $\phi_0$. 

%The DPMD energy is obtained as follows. A dipole spiral of wavelength $N$, with $p_{xy}^s=0.26$~C/m$^2$ and $p_{z}^s=0.58$~C/m$^2$ that are comparable to their values at 300~K, is constructed using a $15\times15\times N$ supercell. The atomic positions are then fully optimized using the \texttt{minimize} procedure as implemented in \texttt{LAMMPS} with the DP model.

\subsection{Strong coupling between neighboring layers }
We can show that any deviation ($\sigma$) from $\delta$ results in an increase in free energy. The total free energy for a dipole spiral with $N>4$ is expressed as:
\begin{equation}
    F=N\left[\mathcal{B}+\mathcal{C}\sin^2\left(\frac{\pi}{N}\right)\right]=N\left[\mathcal{B}+\mathcal{C}\sin^2\left(\frac{\delta}{2}\right)\right],
\end{equation}
where $\mathcal{B}$ represents the local energy contribution and $\mathcal{C} \sin^2\left(\frac{\delta}{2}\right)$ the gradient energy component. If one layer has its $p_{xy}$ rotated by an additional angle $\sigma$,  this rotation does not alter the local energy but does affect the gradient energy. The change in the total free energy is then given by:
\begin{equation}
\begin{aligned}
    \Delta F & = F'-F \\
    &= (N-2)\mathcal{C}\sin^2\left(\frac{\delta}{2}\right) + \mathcal{C}\sin^2\left(\frac{\delta+\sigma}{2}\right) + \mathcal{C}\sin^2\left(\frac{\delta-\sigma}{2}\right)
    - N\mathcal{C}\sin^2\left(\frac{\delta}{2}\right)\\
    &= \mathcal{C} \left[ \sin^2\left(\frac{\delta+\sigma}{2}\right) - \sin^2\left(\frac{\delta}{2}\right) \right] + \mathcal{C} \left[ \sin^2\left(\frac{\delta-\sigma}{2}\right) - \sin^2\left(\frac{\delta}{2}\right) \right]\\
    &= \mathcal{C} \sin \left(\frac{\sigma}{2}\right)\cdot \left[\sin\left(\delta+\frac{\sigma}{2}\right)-\sin\left(\delta-\frac{\sigma}{2}\right)\right] \\
    &= \mathcal{C} \sin^2\left(\frac{\sigma}{2}\right)\cdot\left(2\cos\delta\right) > 0 
    \end{aligned} 
\end{equation}
Since $\Delta F$ is always positive, this implies that the dipoles in neighboring layers tend to maintain the optimal angle $\delta = 2\pi/N$. This tendency helps to explain why the dipole spiral exhibits spontaneous oscillations while maintaining its spiral configuration.

\subsection{Extracting model parameters from DPMD}
\begin{figure}[htbp]
\includegraphics[width=0.98\textwidth]{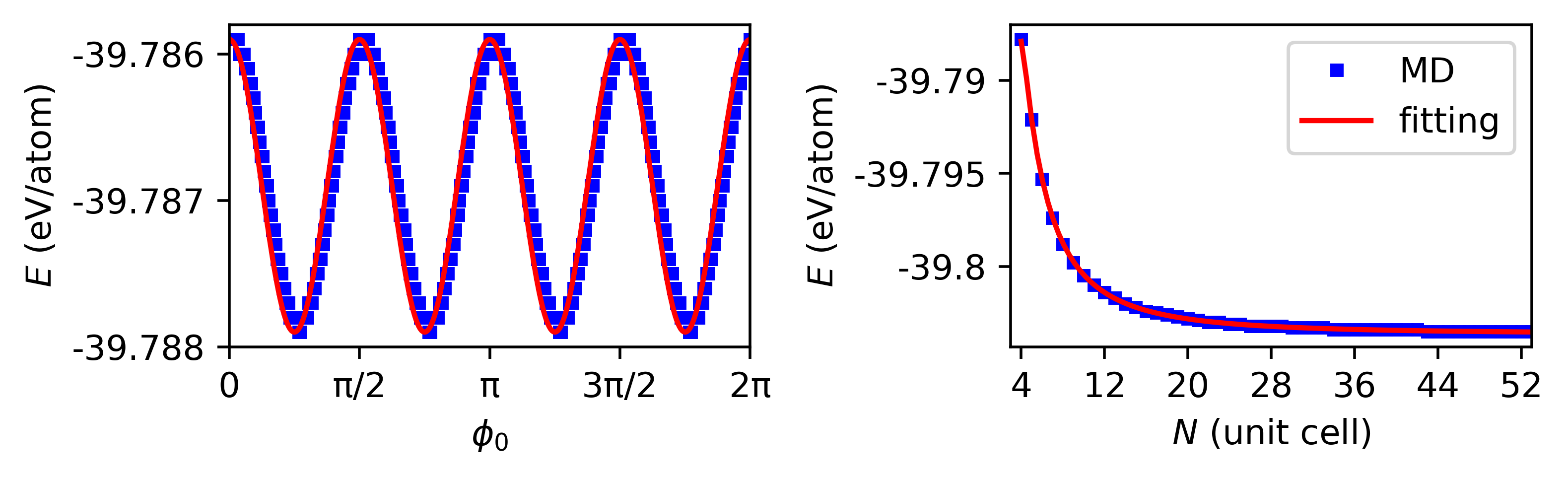}
\caption{(a) Energy per atom as a function of $\phi_0$ for a dipole spiral with $N=4$. (b) Energy evolution with respect to $N$ for dipole spirals. The blue squares represent values obtained using the DP model, and the red lines illustrate the fitted results. 
\label{Fitting}}
\end{figure}

\iffalse
As derived above, the total free energy per layer of a dipole spiral is:
\begin{equation}
\begin{aligned}
    f=f_{\rm loc} + f_g &= \left(\mathcal{A}\frac{1}{N} \sum_{k=1}^N\cos\left(4(\phi_0+k\frac{2\pi}{N})\right) + \mathcal{B}\right)+\mathcal{C}\sin^2\left(\frac{\pi}{N}\right)\\
    &=
    \begin{cases}
       \mathcal{A}\cos(4\phi_0) +\mathcal{B}+\mathcal{C}\sin^2\left(\frac{\pi}{N}\right), & N=1,2,\text{or } 4,\\
       \mathcal{B}+\mathcal{C}\sin^2\left(\frac{\pi}{N}\right), & N=3,\text{or} >4.
    \end{cases}
%    &=\mathcal{A}\cos(4\phi_0) \cdot \mathcal{H}(4-N)+\mathcal{B}\sin^2\left(\frac{\pi}{N}\right)+\mathcal{C}\quad(N\geq4, N\in \mathbb{Z}^+)
    \label{Ftotal}
\end{aligned} 
\end{equation}
\fi
We determine the values of $\mathcal{A}$, $\mathcal{B}$, and $\mathcal{C}$ by fitting equation~(\ref{F4}) to DP energies of dipole spirals with various wavelengths.
Specifically, we model a dipole spiral of wavelength $N$ using a $15\times15\times N$ supercell with  $p_{xy}^s=0.26$~C/m$^2$ and $p_{z}^s=0.58$~C/m$^2$ that are comparable to their values at 300~K. The atomic positions are fully optimized using the \texttt{minimize} procedure as implemented in \texttt{LAMMPS} with the DP model. 
In the case of $N=4$, the energy of a dipole spiral oscillates with respect to $\phi_0$, as shown in Fig.~\ref{Fitting}(a). For $N\ge4$, the energy of the dipole spiral depends solely on $N$. The fitted parameters are $\mathcal{A}=0.001$, $\mathcal{B}=-39.803642$, and $\mathcal{C}=0.033484$.

\newpage

\tikzset{every picture/.style={line width=1.75pt}} %设置默认线宽为1.75pt        
\begin{tikzpicture}[remember picture, overlay]
\node at (current page.west) [anchor=east, xshift=-4.5cm, yshift=11.8cm] {
\begin{tikzpicture}[remember picture, overlay, x=1.05pt,y=1.05pt,yscale=-1,xscale=1]
\node [anchor=base, inner sep=0pt, font=\bfseries\fontsize{14}{16}\selectfont\fontfamily{arial}] at (207,19) {a};
\node [anchor=base, inner sep=0pt, font=\bfseries\fontsize{14}{16}\selectfont\fontfamily{arial}] at (424,19) {b};
\end{tikzpicture}
};
\end{tikzpicture}

\subsection{Entropy contribution to the free energy }
\begin{figure}[htbp]
\includegraphics[width=0.9\textwidth]{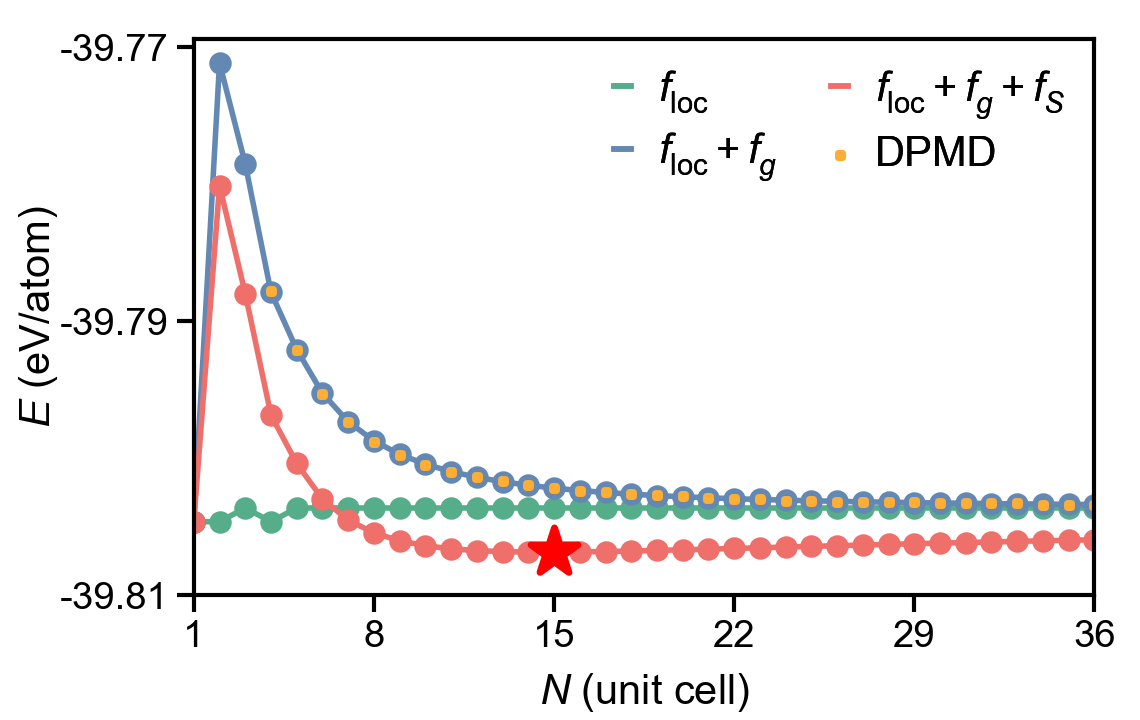}
\caption{Various free energy contributions for a dipole spiral as a function of wavelength $N$.}
\label{FitN}
\end{figure}

For $N>4$, $f_{\rm loc}=\mathcal{B}$ is a constant and $f$ defined by equation~(\ref{F4}) does not dependent on $\phi_0$. This surprising result arises naturally from the identity, $\sum_{k=1}^N \cos\left(4\left(\phi_0 + k\frac{2\pi}{N}\right)\right) = 0$ (see proof in Supplementary IX). 
The rotationally invariant free energy explains the stochastically rotating behavior of the dipole spiral (Fig.~4\textbf{c-d}).
As shown in Fig.~\ref{FitN} , energies of dipole spirals ($f_{\rm loc}+f_g$) is decreasing monotonically with $N$. This suggests that additional effects are necessary to stabilize the spiral at a finite wavelength.  We propose that the dipole spiral is stabilized by entropy, for which we  introduce an additional free-energy term (per layer) that resembles the contribution from Boltzmann entropy, $f_S=\frac{1}{N} k_BT\ln N$, where $k_B$ is the Boltzmann constant and $k_B\ln N$ gives the total entropy.
After introducing an additional free-energy term, $f_S$, that resembles the contribution from Boltzmann entropy, the total free energy per layer for a dipole spiral with $N>4$ is:
\begin{equation}
\begin{aligned}
    f&=f_{\rm loc} + f_g + f_S\\
    &=\mathcal{B}+\mathcal{C}\sin^2\left(\frac{\pi}{N}\right)-\frac{1}{N} k_BT\ln N.
    \label{FtotalS}
\end{aligned} 
\end{equation}
We find that $f$ reaches the minimum at $N=15$, as shown in Fig.~\ref{FitN}.

\clearpage
\newpage

\section{Additional DFT and MD modeling of dipole spirals}

\subsection{DFT modeling of dipole spirals}
\begin{figure}[htbp]
\centering
\includegraphics[width=0.3\textwidth]{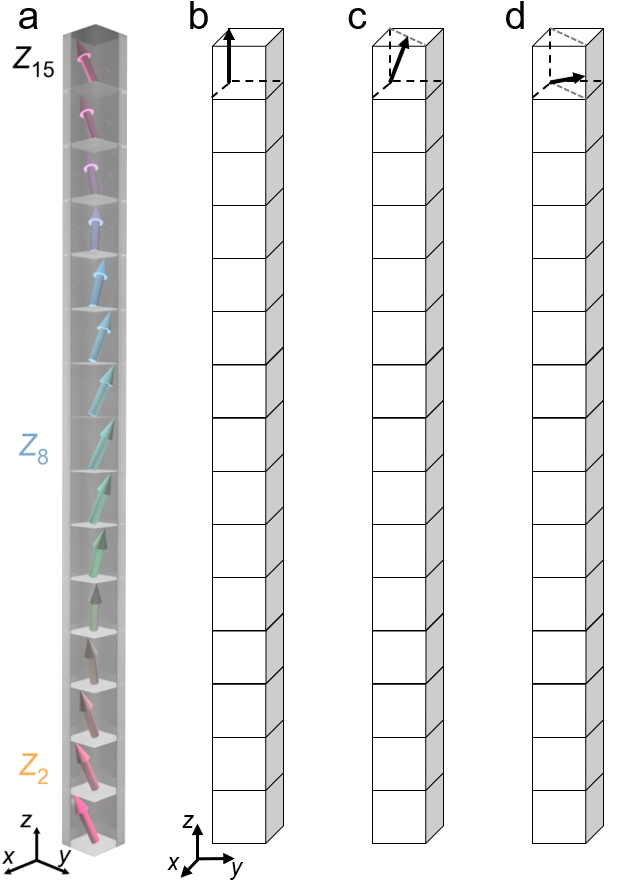}
\caption{Schematics of a dipole spiral, singe-domain [001], $M_A$ and $M_B$ states modeled with  $1\times1\times15$ supercells in DFT.}
\label{schematic}
\end{figure}

We compute the DFT energies of  $1\times1\times15$ supercells in four different polar states: (1) a dipole spiral with dipoles rotating progressively around the $z$-axis, (2) a single-domain [001] state with all unit cells having polarization aligned along [001], (3) a single-domain $M_A$ $[uu1]$ state, and (4) a single-domain $M_B$ $[11u]$ state, as depicted in Fig.~\ref{schematic}. The in-plane lattice constants are fixed at $a_{\mathrm{IP}}= b_{\mathrm{IP}}=3.948$~\AA, while the out-of-plane lattice constant and atomic positions aer fully relaxed.
As shown in Table~\ref{DFTenergy}, the dipole spiral is lower in energy compared to the other three single-domain states, further corroborating results from MD simulations. It is noteworthy that the DP model also correctly predicts the dipole spiral state to be lower in energy than the single-domain [001] state by 12.1 meV.

\vspace{-0.12in}
\begin{table}[htbp]
\setlength{\belowcaptionskip}{0.05 in}
\caption{DFT absolute energies ($E$ in eV) and relative energies ($\Delta E$ in meV) of four different polar states computed with $1\times1\times15$ supercells. The single-domain [001] state is chosen as the reference for the calculations of $\Delta E$.}
\centering
\begin{tabular}
{c|cccc}
\hline
\hline
  & Dipole Spiral   & [001] & $M_A$ $[uu1]$ & $M_B$ $[11u]$\\
\hline
$E$ (eV) &  $-597.231285$ & $-597.221633$ & $-597.193492$ & $-597.184068$ \\
%E(DP) (eV)   &  -597.249980 & -597.237870 & -597.232260 & -597.226080 \\
$\Delta E$ (meV) & $-9.7$ & 0 &28.1 &37.6 \\
%E(DP) relative to [001] state (eV) & -0.012110 & 0 & 0.005610 & 0.011790 \\

\hline
\hline
\end{tabular}
\label{DFTenergy}
%\caption{DFT energies of 1×1×15 supercells adopting four different polar states}
\end{table}

%\vspace{-0.2in}

\newpage
\subsection{Temperature-driven evolution of polarization and $d_{33}$ for a dipole spiral}
\begin{figure}[htbp]
\centering
\includegraphics[width=0.6\textwidth]{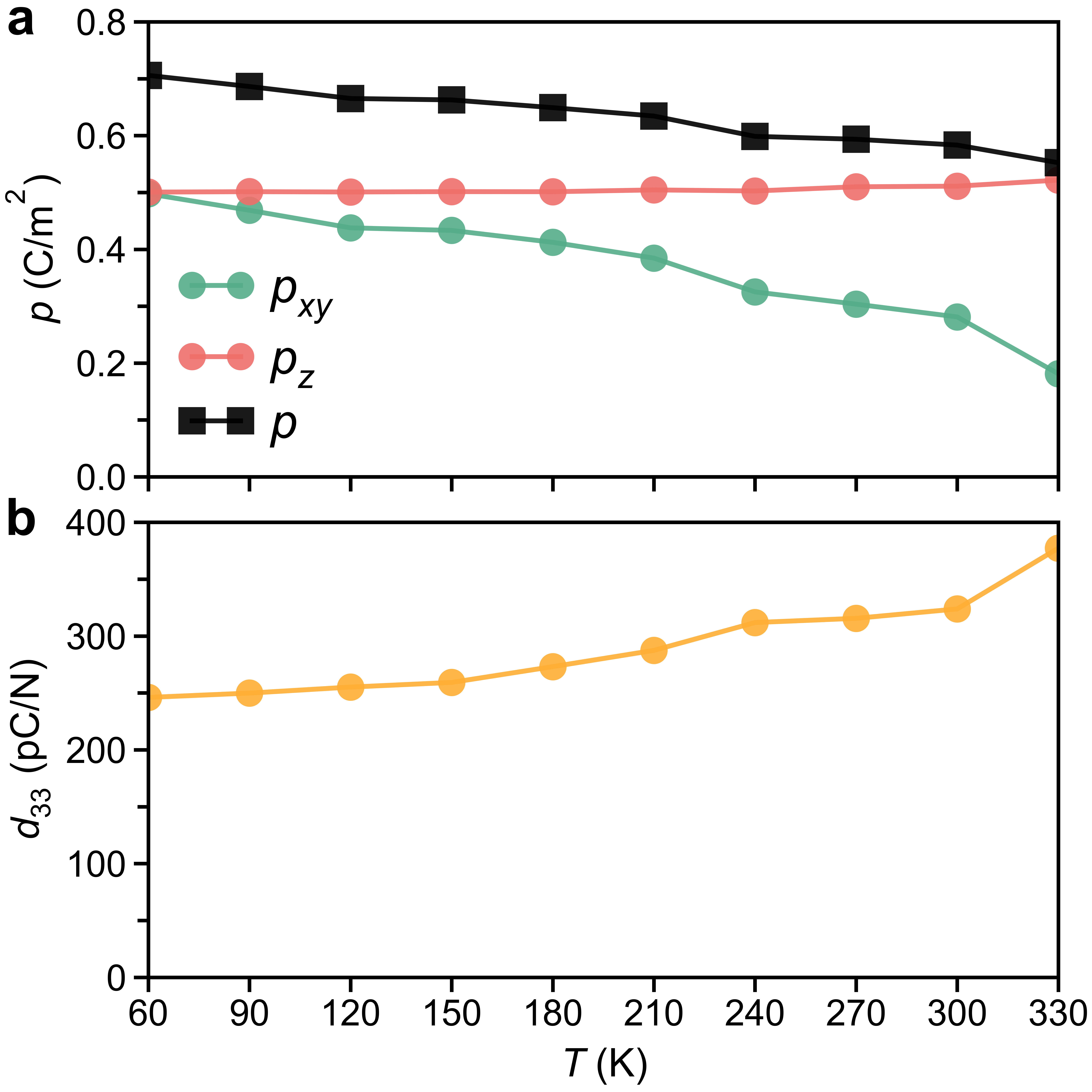}
\caption{Temperature-dependent polarization values and piezoelectric coefficient $d_{33}$.
(a) Evolution of layer-resolved in-plane polarization $p_{xy}$, out-of-plane polarization $p_z$, and the total polarization $p$ as the temperature ($T$) increases. (b) Variation of the piezoelectric coefficient $d_{33}$ as a function of $T$.
}
\label{spiralvsT}
\end{figure}

We perform MD simulations at a strain state defined by $a_{\mathrm{IP}}=b_{\mathrm{IP}}=3.958$~\AA, with temperature ranging from 60 to 330~K. 
Figure~\ref{spiralvsT}(a) illustrates the temperature-driven evolution of layer-resolved in-plane polarization $p_{xy}$, out-of-plane polarization $p_z$, and the total polarization $p$ for a dipole spiral with a wavelength of 15 unit cells. Interestingly, the magnitude of $p_z$ remains nearly constant, while both $p_{xy}$ and $p$ decrease with increasing temperature. At a specific temperature, the value of $d_{33}$ is derived from the slope of the strain-electric field curve, which is obtained by applying varying electric fields along the $z$-axis in MD simulations. We find that $d_{33}$ is enhanced at higher temperatures, although its thermal sensitivity is moderate, as shown in Fig.~\ref{spiralvsT}(b).
The average value of $d_{33}$ is $\approx 290$~pC/N across a temperature range of 250 K, indicating a temperature-stable, large piezoelectric response over a broad range of operational temperatures.

\newpage

\subsection{Strain-driven evolution of polarization for a dipole spiral at 300~K}

We examine the evolution of layer-resolved in-plane polarization ($p_{xy}$) and out-of-plane polarization ($p_z$) as a function of the in-plane lattice constant $a_{\rm IP}$, which ranged from 3.952 to 3.956 \AA~with a fine resolution of 0.0004 \AA~in the changes of $a_{\rm IP}$. The supercell adopts a single $c$-domain state at $a_{\rm IP}=3.952$ \AA~and transitions to a dipole spiral at $a_{\rm IP}=3.956$~\AA.

\begin{figure}[htbp]
\centering
\includegraphics[width=0.98\textwidth]{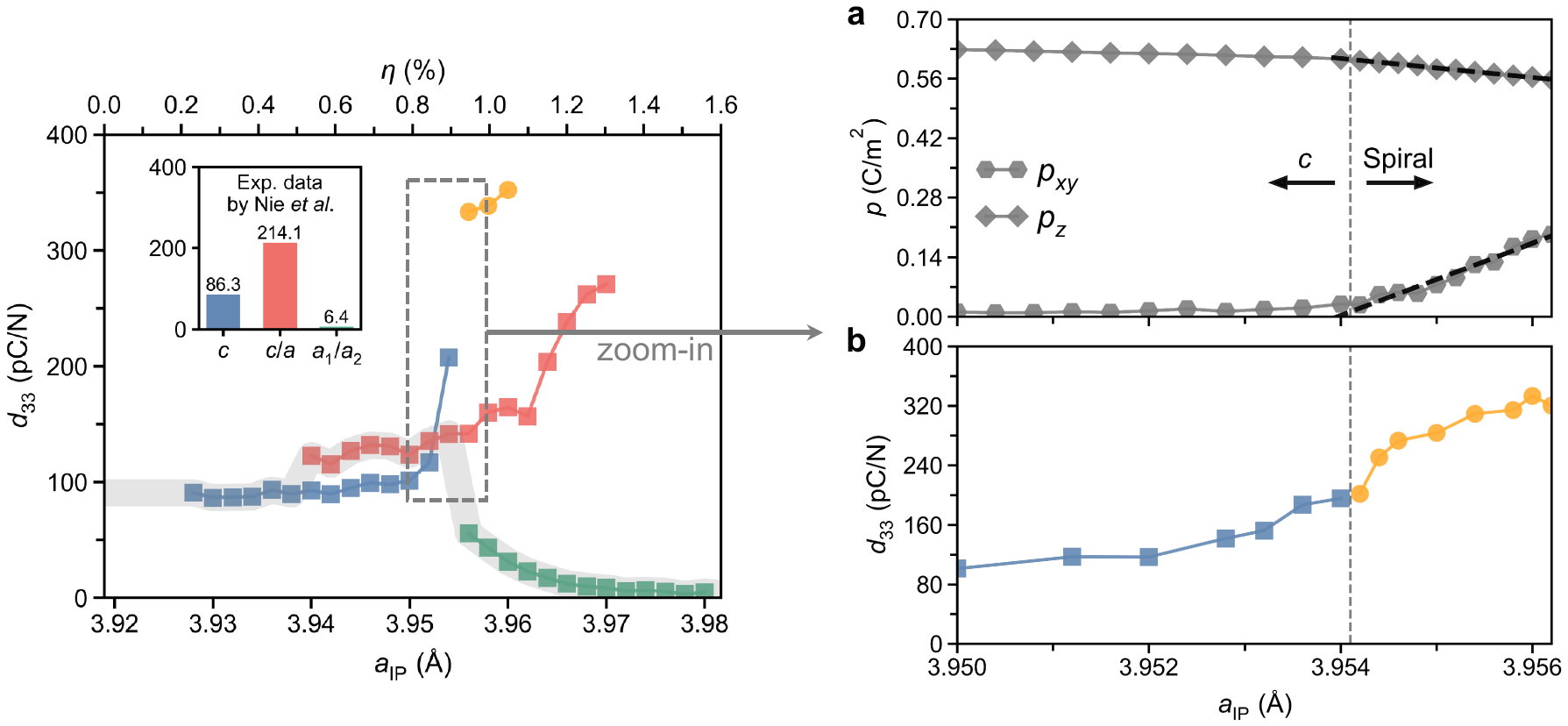}
\caption{Evolution of (a) layer-resolved in-plane polarization ($p_{xy}$) and out-of-plane polarization ($p_{z}$) and (b) piezoelectric coefficient ($d_{33}$) with respect to in-plane lattice constants ($a_{\rm IP}$) obtained with MD simulations at 300~K.}
\label{c2spiral}
\end{figure}

As shown in Fig.~\ref{c2spiral}, the overall trend is that the magnitude of $p_z$ decreases with increasing tensile strain, which corresponds with an increase in the magnitude of $p_{xy}$. Although the curves for polarization evolution appear smooth, we identified a discontinuity at a critical in-plane lattice constant of $a_{\rm IP}^{\rm tr}=3.9541$~\AA. Below this critical value, the changes in $p_{xy}$ and $p_z$ with respect to $a_{\rm IP}$ are minimal, exhibiting only slight slopes. Above $a_{\rm IP}^{\rm tr}$, however, both $p_z$ and $p_{xy}$ exhibit more pronounced changes. Simultaneously, the value of $d_{33}$ also shows a sharper increase once $a_{\rm IP}$ surpasses $a_{\rm IP}^{\rm tr}$.

\subsection{Map temperatures in MD to experimental temperatures}
\begin{figure}[htp]
\centering
\includegraphics[width=0.6\textwidth]{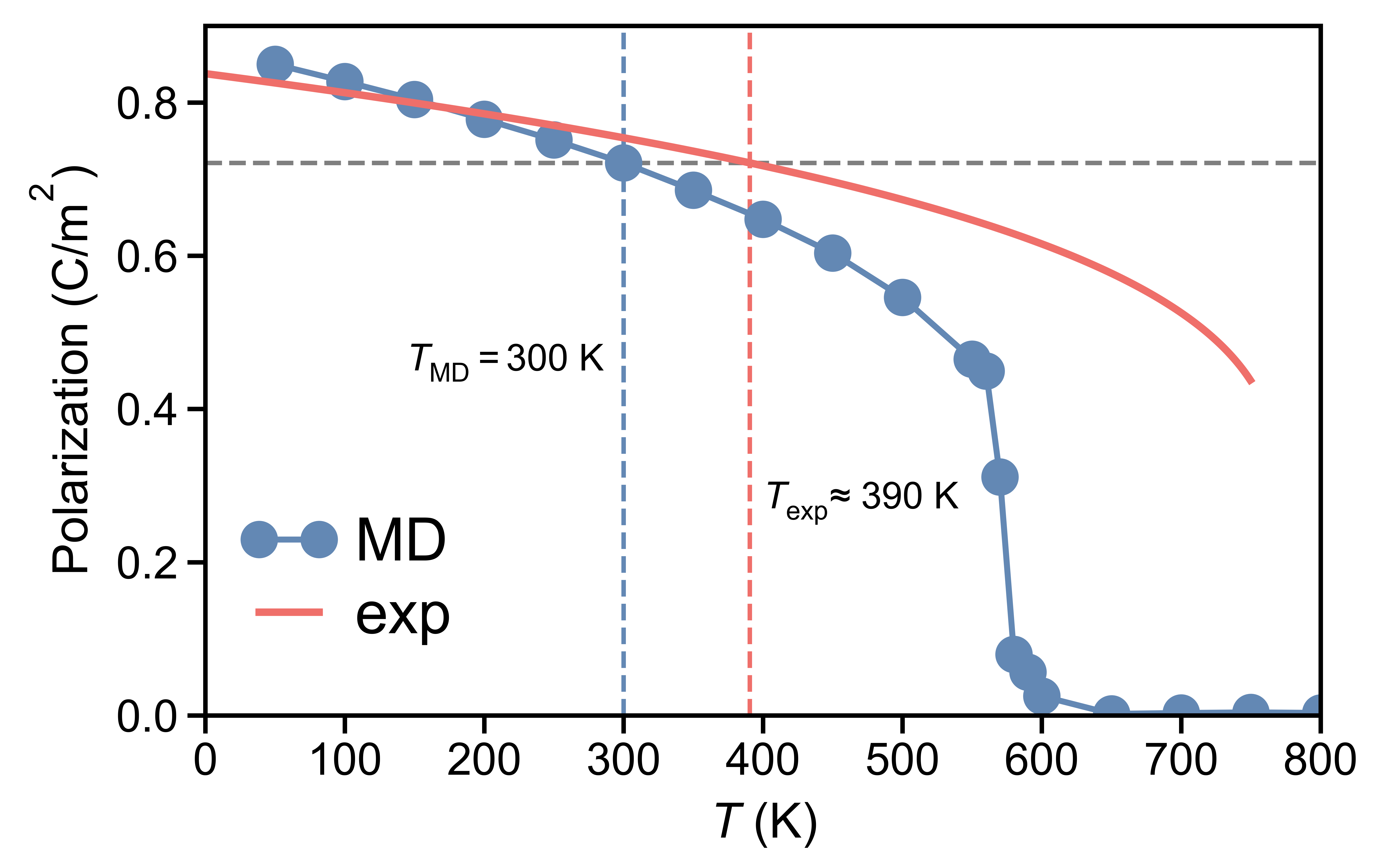}
\caption{Polarization as a function of temperature. The experimental data is taken from ref.~\cite{Haun87p3331}. For a simulated temperature ($T_{\rm MD}$) in MD, its corresponding experimental temperature ($T_{\rm exp}$) will result in the same polarization magnitude. 
}
\label{Tshift}
\end{figure}

The ferroelectric-paraelectric  phase transition temperatures of Pb$_x$Sr$_{1-x}$TiO$_3$ solid solutions, as predicted by the DP model from MD simulations, are generally lower than experimental values. 
We developed a protocol to map the temperatures in MD simulations ($T_{\rm MD}$) to experimental temperatures ($T_{\rm exp}$).
First, we extracted experimental data on the temperature-dependence of polarization ($P$) from ref.~\cite{Haun87p3331}, represented by the red solid line in Fig.~\ref{Tshift}.
Since the DP model accurately reproduces the ground-state polarization value (at 0 K), we establish a straightforward relationship between $T_{\rm MD}$ and $T_{\rm exp}$: the value of $T_{\rm exp}$ is determined by matching the polarization at $T_{\rm MD}$. As shown in Fig.~\ref{Tshift}, a $T_{\rm MD}$ of 300 K corresponds approximately to a $T_{\rm exp}$ of 390 K, while a $T_{\rm exp}$ of 300 K is roughly equivalent to 210 K in MD simulations. 

%\vspace{-0.2in}

\subsection{Switch 90$^\circ$ walls with $P_x$ components}

\begin{figure}[htbp]
\centering
\includegraphics[width=0.98\textwidth]{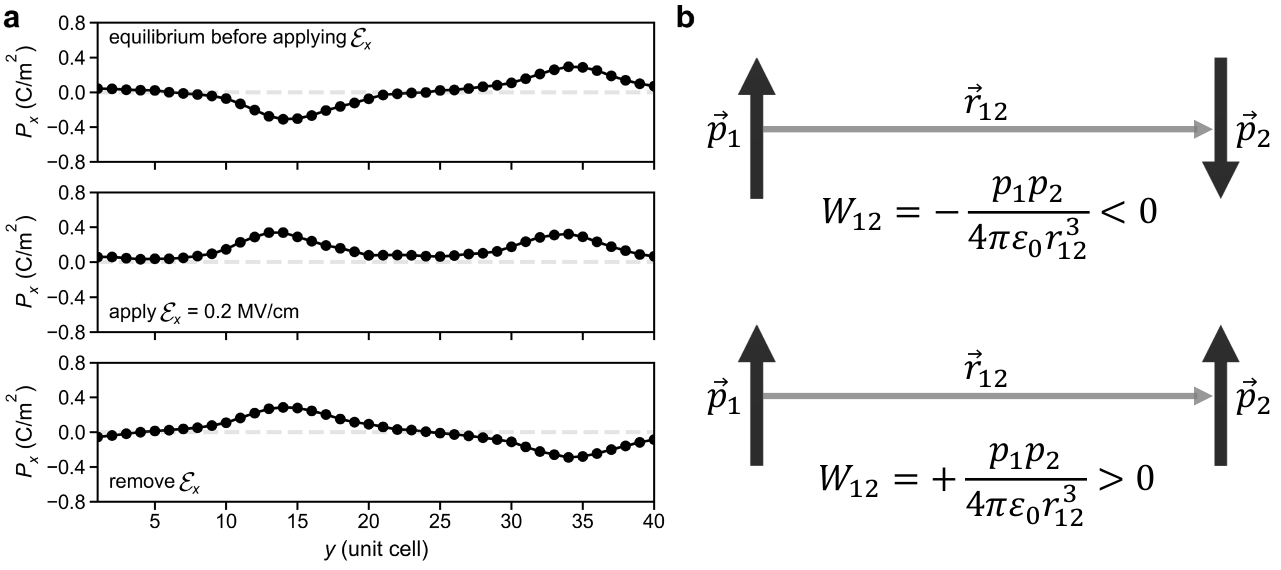}
\caption{
(a) Electric-field-driven response of 90$^\circ$ domain walls  with $P_x$ components separating $-P_y$ and $+P_z$ domains. The strain state is the same as that reported in Fig.~3\textbf{f} of the main text. An electric field $\mathcal{E}_x$ applied along the $x$-direction can switch the $-P_x$ wall. After the removal of $\mathcal{E}_x$, neighboring walls return to the state characterized by anti-parallel $P_x$ components. (b) Schematic illustrating the electrostatic energy for parallel and antiparallel dipoles.  
}
\label{Px}
\end{figure}

As illustrated in Fig.~3\textbf{f}, domain walls separating $-P_y$ and $P_z$ domains exhibit $\pm P_x$ components, while adopting anti-parallel coupling between adjacent walls. 
Our finite-field MD simulations revealed that when an electric field ($\mathcal{E}_x$) is applied along the $+x$ direction to a $c/a$ two-domain state containing 90$^\circ$ walls with $\pm P_x$ components, the $-P_x$ component flips to $+P_x$, as shown in in Fig.~\ref{Px}~(a). However, upon removal of the electric field, the domain structure reverts to a state with anti-parallel $P_x$ components. This behavior strongly suggests a preference for adjacent domain walls to maintain anti-parallel alignment of $P_x$ components.

Interestingly, during the relaxation process after the removal of the electric field, which wall will flips its $+P_x$ is probabilistic. 
We observed that the domain wall initially characterized by a $-P_x$ component before applying $\mathcal{E}_x$ stays at $+P_x$, while the opposing wall transitions to the $-P_x$ state. These findings indicate that the states $+P_x$ and $-P_x$ are energetically equivalent within each wall. However, there is a strong thermodynamic preference for anti-parallel coupling between neighboring walls, likely due to long-range Coulomb interactions. As illustrated in Fig.~\ref{Px} (b), dipoles aligned antiparallel along the direction perpendicular to the dipoles result in lower electrostatic energy compared to parallel alignment.

\subsection{Domain wall thickness in $c/a$ two-domain states}

\begin{figure}[htbp]
\centering
\includegraphics[width=0.98\textwidth]{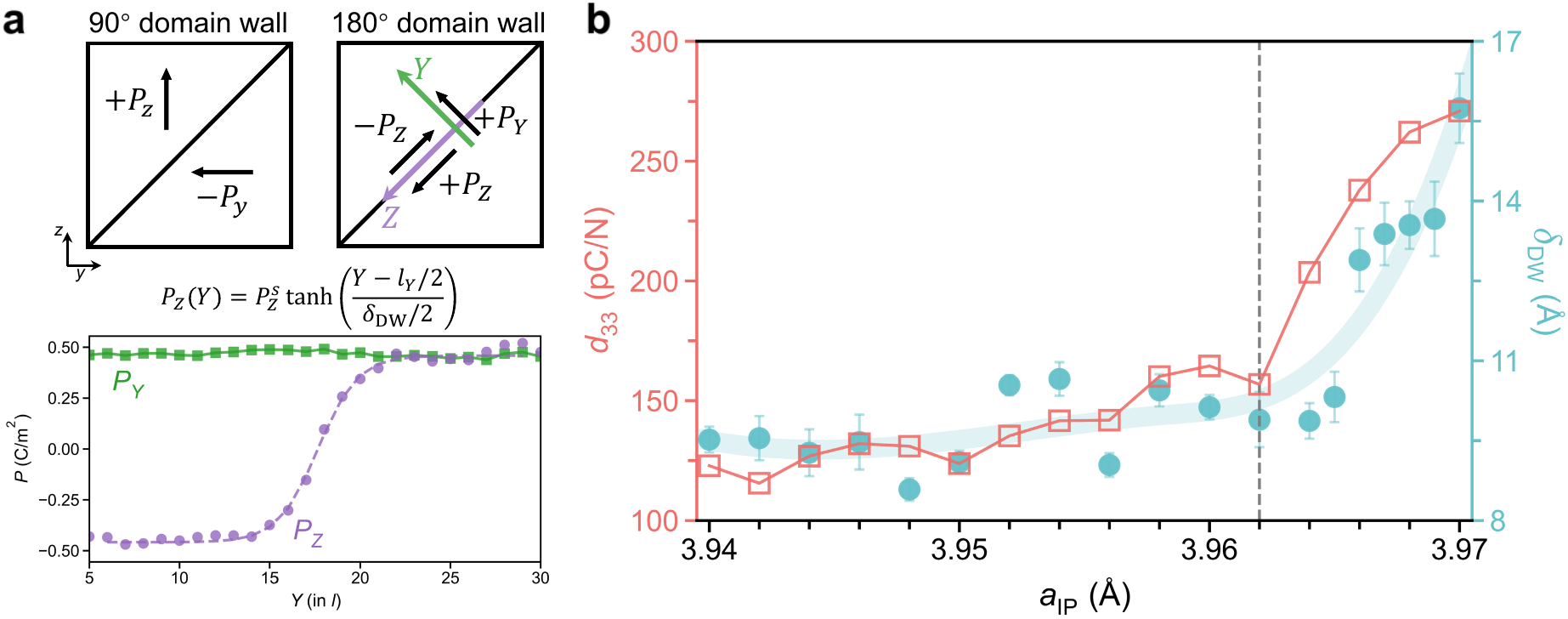}
\caption{
(a) Schematic of
mapping a 90$^\circ$ domain wall in $y$-$z$ coordinates to a 180$^\circ$  domain wall in
$Y$–$Z$ coordinates. 
The bottom panel shows the polarization profile of a
90$^\circ$ domain wall in $Y$-$Z$ coordinates; $l$ denotes the spacing between neighboring Ti lattice planes along the $Y$-axis, which is approximately $(a+c)/2\sqrt{2}$.
(b) $d_{33}$ and $\delta_{\rm DW}$ as a function of the in-plane lattice constant ($a_{\rm IP}$) at 300~K.}
\label{DWfit1}
\end{figure}

To quantify the thickness of a 90$^\circ$ domain wall separating $-P_y$ and $+P_z$ domains (as reported in Fig.~\textbf{3} of the main text), we performed a coordinate transformation as illustrated in Fig.~\ref{DWfit1}~(a). A 90$^\circ$ domain wall in $y$–$z$ coordinates
can be viewed as a special 180$^\circ$ domain wall in $Y$-$Z$ coordinates: the
polarization component parallel to the wall ($P_Z$) is reversed by 180$^\circ$ across the boundary, while the component perpendicular to the wall ($P_Y$) remains nearly unchanged. We fitted the $P_Z$ profile using $P_Z(Y) = P_Z^s\tanh(\frac{Y-l_Y/2}{\delta_{\rm DW}/2})$, where $\delta_{\rm DW}$ represents the domain wall thickness. 

We then quantified $\delta_{\rm DW}$ as a function of the in-plane lattice constant ($a_{\rm IP}$). For a specific strain, we analyzed 5 instantaneous polarization profiles of $c/a$ two-domain states and averaged the fitted $\delta_{\rm DW}$ values. As shown in Fig.~\ref{DWfit1}~(b), we observed a general increase in $\delta_{\rm DW}$ with increasing $a_{\rm IP}$.  Notably, the rapid increase in $\delta_{\rm DW}$ beyond a critical tensile strain of $a_{\rm IP}=3.962$~\AA~coincides with a rapid rise in $d_{33}$ and the emergence of a significant polarization component ($P_x$) within domain walls. The application of an external field changes the ratio of the volumes of the $c$ and $a$ domains, which is responsible for the overall strain change. This change is due to the collective and coordinated small-angle rotations of dipoles at the domain walls, analogous to ``coordinated gear dynamics." A domain wall with a broader thickness also suggests lower rotational barriers for dipoles near the domain wall.  

\clearpage
\newpage
\section{Superlattices supporting dipole spiral arrays}
\begin{figure}[htbp]
\centering
\includegraphics[width=0.76\textwidth]{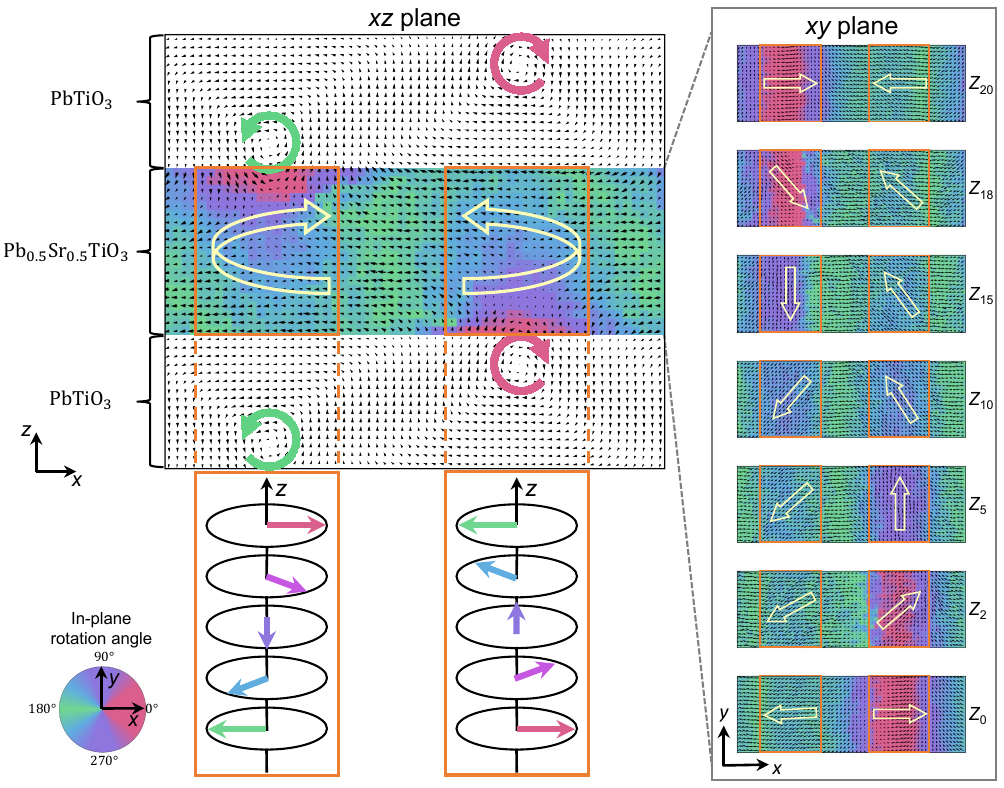}
\caption{Dipole spiral arrays in (PbTiO$_3$)$_{16}$/(Pb$_{0.5}$Sr$_{0.5}$TiO$_3$)$_{20}$ superlattices. A $60\times20\times36$ supercell of 216,000 atoms is used in MD simulations at 300~K. Arrows represent the local electric dipoles within each unit cell. The arrows in Pb$_{0.5}$Sr$_{0.5}$TiO$_3$ layers are scaled up by a factor of 2 for better visualization, and the background is colored based on the in-plane rotation angle. Within each spiral, the in-plane dipoles exhibit a 180$^\circ$ rotation from bottom to top. It is possible to further induce out-of-plane polarization component by fine tuning the composition and strain.}
\label{superlattice}
\end{figure}

We have designed all-ferroelectric superlattices composed of alternating layers of PbTiO$_3$ and Pb$_{0.5}$Sr$_{0.5}$TiO$_3$.
Compared to the well-known PbTiO$_3$/SrTiO$_3$ superlattices that support a rich spectrum of ferroelectric topological structures, 
substituting nonpolar SrTiO$_3$ with ferroelectric Pb$_{0.5}$Sr$_{0.5}$TiO$_3$ introduces in-plane ferroelectric polarization. This modification likely helps to alleviate the polarization/dielectric discontinuity at the interface and reduce the depolarization effects. As depicted in Fig.~\ref{superlattice}, this layered heterostructure hosts arrays of dipole spirals in Pb$_{0.5}$Sr$_{0.5}$TiO$_3$ layers, each linking a pair of polar vortices within PbTiO$_3$ layers. These findings underscore the potential of utilizing advanced thin-film deposition techniques to experimentally realize the dipole spiral proposed in this study in a realistic setting. 

\clearpage
\newpage
\section*{Appendix}\label{Mathematical proof}

%\noindent \textbf{Proposition:} $\sum_{k=1}^{N}\cos\left(4(\phi_0+2\pi k/N)\right)=0$ $(N>4, N\in \mathbb{Z} )$\\

\noindent Proof $\sum_{k=1}^{N}\cos\left(4(\phi_0+2\pi k/N)\right)=0$ $(N>4, N\in \mathbb{Z} ).$\\
For convenience, we change the summation to run from $k=0$ to $k=N-1$:
\[\sum_{k=1}^{N}\cos\left(4(\phi_0+2\pi k/N)\right)=\sum_{k=0}^{N-1}\cos\left(4(\phi_0+2\pi k/N+2\pi/N)\right) = \sum_{k=0}^{N-1}\cos\left(4(\phi_0'+2\pi k/N)\right) \]
Using the compound angle formula, we obtain:
\begin{equation}
    \sum_{k=0}^{N-1}\cos(4(\phi_0'+2\pi k/N))=\sum_{k=0}^{N-1}\left[\cos(8\pi k/N)\cos(4\phi_0')-\sin(8\pi k/N)\sin(4\phi_0')\right]
    \label{cos}
\end{equation}
Note that the roots of $x^N-1=0$ are:
\[
e^{i {\frac{8\pi\cdot0}{N}}}, e^{i {\frac{8\pi\cdot1}{N}}}, e^{i {\frac{8\pi\cdot2}{N}}},..., e^{i {\frac{8\pi\cdot(N-1)}{N}}}
\]
According to Vieta's formulas which relate the polynomial coefficients to signed sums of products of the roots, it follows that:
\begin{equation}
    \sum_{k=0}^{N-1}e^{i {\frac{8\pi k}{N}}}=0
    \label{plus}
\end{equation}
Similarly, it is easy to show:
\begin{equation}
    \sum_{k=0}^{N-1}e^{-i {\frac{8\pi k}{N}}}=0
    \label{minus}
\end{equation}
The sum of equations~(\ref{plus}) and (\ref{minus}) yields:
\begin{equation}
    0=\sum_{k=0}^{N-1}\left(e^{i {\frac{8\pi k}{N}}} + e^{-i {\frac{8\pi k}{N}}} \right)
    =\sum_{k=0}^{N-1}2\cos(8\pi k/N),
    \label{plus2}
\end{equation}
while the difference between equations~(\ref{plus}) and (\ref{minus}) gives:
\begin{equation}
    0=\sum_{k=0}^{N-1}\left(e^{i {\frac{8\pi k}{N}}} - e^{-i {\frac{8\pi k}{N}}} \right)
    =\sum_{k=0}^{N-1}2i\sin(8\pi k/N).
    \label{minus2}
\end{equation}
The substitution of equations~(\ref{plus2}) and (\ref{minus2}) into equation~(\ref{cos}) proves
\begin{equation}
    \sum_{k=1}^{N}\cos\left(4(\phi_0+2\pi k/N)\right)=\sum_{k=0}^{N-1}\cos(4(\phi_0'+2\pi k/N))=0
\end{equation}

\bibliography{SL.bib}